\documentclass[journal]{IEEEtran}

\usepackage{graphicx}
\def\BibTeX{{\rm B\kern-.05em{\sc i\kern-.025em b}\kern-.08em
    T\kern-.1667em\lower.7ex\hbox{E}\kern-.125emX}}
\usepackage[font=small,labelfont=bf,tableposition=top,hypcap=false]{caption}
\usepackage{blindtext}

\let\oldtwocolumn\twocolumn
\renewcommand\twocolumn[1][]{%
    \oldtwocolumn[{#1}{
    \begin{center}
    \vskip-5ex
        \centering
        \begin{tikzpicture}
\node[anchor=south west,inner sep=0] (overview) at (0,0)
  {\includegraphics[width=\textwidth]{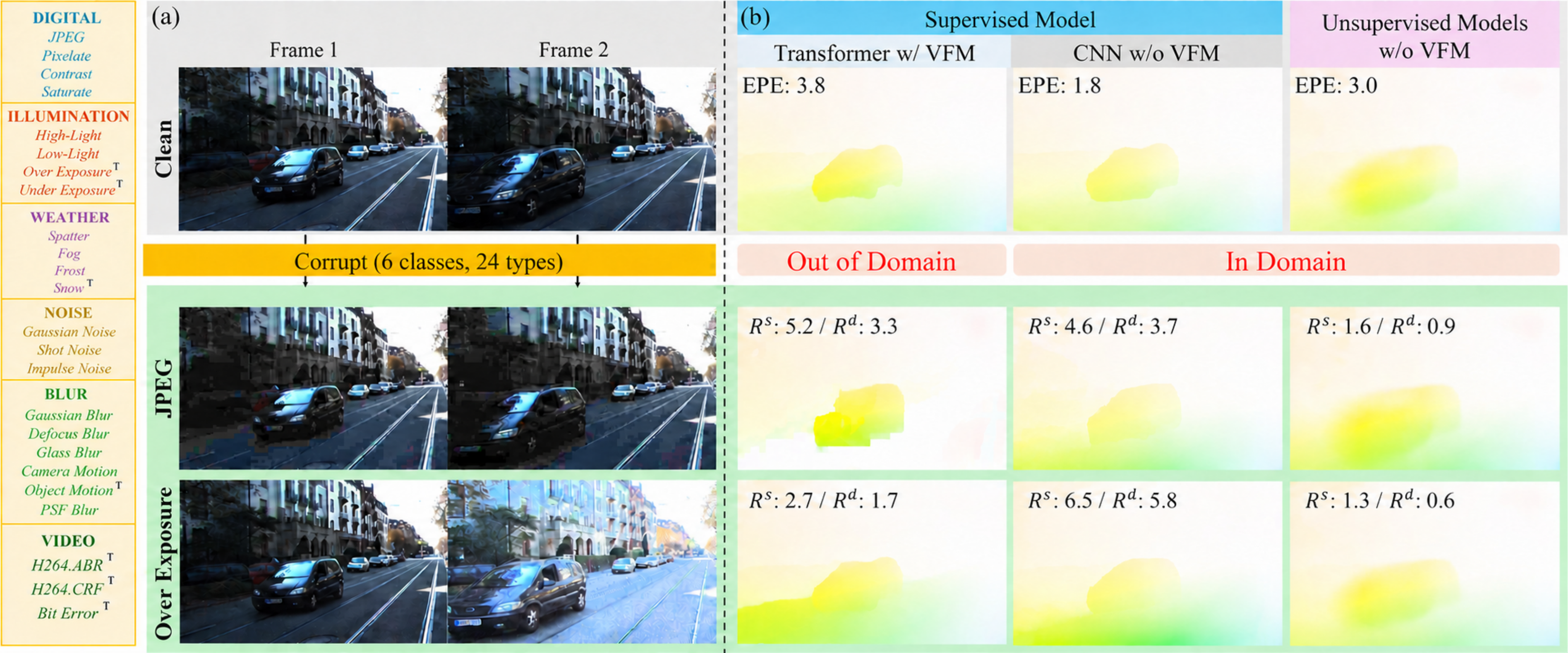}};
\begin{scope}[x={(overview.south east)},y={(overview.north west)}]
\foreach \x/\method in {0.557/SAMFlow-H,0.733/RAFT,0.910/ARFlow} {
  \foreach \y in {0.655,0.290,0.020} {
    \node[anchor=south,
      inner sep=1.2pt,text=black,
      font=\fontfamily{ptm}\fontseries{m}\fontshape{n}\scriptsize]
      at (\x,\y) {\method};
  }
}
\end{scope}
\end{tikzpicture}

        \captionof{figure} {{\textbf{Overview of the robustness evaluation framework for optical flow estimation.}
        (a) Controlled spatial, asymmetric inter-frame, and video-pipeline corruptions are applied to captured source data to construct \textbf{KITTI Flow Corruptions (KITTI-FC)} for driving-domain robustness and \textbf{GoPro Flow Corruptions (GoPro-FC)} for motion-distribution robustness. The vertical list denotes the 24 corruption types organized into six categories. The superscript $T$ marks the seven temporal corruptions that explicitly alter inter-frame consistency.
        (b) We evaluate optical-flow model variants under Out-of-Domain (OOD) and In-Domain (ID) settings using complementary output-stability error $R^{s}$ and accuracy-degradation $R^{d}$. Prediction columns (left to right): SAMFlow-H~\cite{zhou2024samflow, kirillov2023segment}, RAFT~\cite{teed2020raft}, and ARFlow~\cite{liu2020learning}. VFM denotes vision foundation model.
        }}
        \label{fig:overview}
    \end{center}
    }]
}

\usepackage{amsmath,amsfonts,amssymb}
\usepackage{algorithmic}
\usepackage{algorithm}
\usepackage{array}
\usepackage{booktabs,tabularx}
\usepackage[caption=false,font=normalsize,labelfont=sf,textfont=sf]{subfig}
\usepackage{textcomp}
\usepackage{stfloats}
\usepackage{url}
\usepackage{verbatim}
\usepackage{cite}
\usepackage{placeins}

\usepackage{amsmath}
\usepackage{multirow}
\usepackage{threeparttable}
\usepackage{colortbl}
\usepackage{makecell}
\usepackage{tikz}
\usepackage{tkz-kiviat}
\usepackage{paralist}

\usepackage{utfsym}

\usepackage{xcolor}
\definecolor{rblue}{rgb}{0,0.5,1}
\definecolor{hollywoodcerise}{rgb}{0.96, 0.0, 0.63}
\definecolor{lasallegreen}{rgb}{0.03, 0.47, 0.19}
\definecolor{hanpurple}{rgb}{0.32, 0.09, 0.98}
\definecolor{green(pigment)}{rgb}{0.0, 0.65, 0.31}

\usepackage[pagebackref=false,breaklinks=true,colorlinks,bookmarks=false]{hyperref}
\hypersetup{colorlinks,linkcolor={red},citecolor={hanpurple},urlcolor={magenta}}

\begin{document}

\title{Towards Dependable Optical Flow: Benchmarks for Real-World Spatiotemporal Corruptions}

\author{Zhonghua Yi$^{1,}$, Hao Shi$^{4,1}$, Qi Jiang$^1$, Yao Gao$^1$, Ze Wang$^1$, Yufan Zhang$^1$,\\Kailun Yang$^{2,3,}$\IEEEauthorrefmark{2}, and Kaiwei Wang$^{1,5,}$\IEEEauthorrefmark{2}%
\thanks{This work was supported in part by the Natural Science Foundation of Zhejiang Province (Grant No. LZ24F050003), in part by the National Natural Science Foundation of China (NSFC) under Grant No. 12174341 and No. 62473139, in part by the Hunan Provincial Research and Development Project (Grant No. 2025QK3019), and in part by the State Key Laboratory of Autonomous Intelligent Unmanned Systems (Project No. ZZKF2025-2-10).}
\thanks{$^1$Z. Yi, H. Shi, Q. Jiang, Y. Gao, Z. Wang, Y. Zhang, and K. Wang are with the National Research Center for Optical Instrumentation, Zhejiang University, Hangzhou 310027, China (email: yizhonghua@zju.edu.cn; haoshi@zju.edu.cn; qijiang@zju.edu.cn; gaoyao@zju.edu.cn; 22030039@zju.edu.cn; zhangyufan@zju.edu.cn; wangkaiwei@zju.edu.cn).}%
\thanks{$^2$K. Yang is with the School of Artificial Intelligence and Robotics, Hunan University, Changsha 410012, China (email: kailun.yang@hnu.edu.cn).}%
\thanks{$^3$K. Yang is also with the National Engineering Research Center of Robot Visual Perception and Control Technology, Hunan University, Changsha 410082, China.}%
\thanks{$^4$H. Shi is also with the Ant Group Company Ltd., Hangzhou 310013, China (email: sh531467@antgroup.com).}%
\thanks{$^5$K. Wang is also with the Central Research Institute of Sunny Optical Technology, Hangzhou 315400, China}%
\thanks{\IEEEauthorrefmark{2}Corresponding authors: Kaiwei Wang and Kailun Yang.}%
}

\markboth{IEEE Transactions on Circuits and Systems for Video Technology, October~2026}%
{Yi~\MakeLowercase{\textit{et al.}}: Benchmarking the Robustness of Optical Flow Estimation to Common Corruptions}

\maketitle

\begin{abstract}
Optical flow estimation is a fundamental task in dynamic scene understanding and computer vision. While recent deep models achieve impressive accuracy on clean benchmarks, their robustness to common visual and temporal degradations remains insufficiently explored. To address this gap, we establish a real-video-based spatiotemporal corruption benchmark suite that complements rendered and mixed-source robustness benchmarks, systematically evaluating two complementary aspects of visual perception: domain robustness and motion-distribution robustness. We construct two targeted benchmarks: KITTI Flow Corruptions (KITTI-FC) for dynamic outdoor scenes and GoPro Flow Corruptions (GoPro-FC) for diverse first-person-view complex motions. Both benchmarks use captured source data and apply 24 controlled, predominantly algorithmic corruptions encompassing digital, illumination, weather, noise, blur, and video artifacts. To quantify robustness, we use two complementary axes: the GT-free output-stability error $R^{s}$ and the GT-based accuracy degradation $R^{d}$, together with their clean-error-normalized variants. We systematically evaluate 33 model variants from 17 representative optical flow methods. Our results show that, under the evaluated corruption schedules, models exhibit greater vulnerability to local structural information loss (\textit{e.g.}, JPEG compression and spatter) than to global perceptual distortions (\textit{e.g.}, fog and saturation). We further find that in-domain fine-tuning does not guarantee relative robustness and that no single current method dominates across all tested source and motion settings. Our work provides a controlled diagnostic framework for assessing optical flow robustness and developing reliable perception systems in complex environments. Code and benchmark resources are publicly available at \url{https://github.com/ZhonghuaYi/optical_flow_robustness_benchmark}.
\end{abstract}


\section{Introduction}

Optical flow estimation, which computes dense motion fields between frames, is a fundamental perception task for computer vision and video analysis~\cite{sanchez2011differential,allaert2022comparative,zhang2022optical,lu2024optical,du2024optical}. It provides critical motion cues for essential functions such as object tracking and visual odometry~\cite{capito2020optical, wang2021end, qin2024rso,sormoli2024optical,wang2025enhancing,senst2012robust}. The reliability of these downstream tasks, and ultimately the safety of the entire system, hinges on the accuracy and robustness of the underlying optical flow. While recent deep learning models~\cite{teed2020raft, huang2022flowformer, sea_raft} have demonstrated impressive performance on standard benchmarks, their resilience to common visual and temporal degradations on captured scenes remains a critical, yet underexplored, vulnerability. Minor degradations from adverse weather, sensor noise, or image blur can lead to substantial failures in motion estimation, posing a threat to system reliability.

While robustness evaluation is well established for static image tasks~\cite{hendrycks2019benchmarking,dong2023benchmarking}, systematically assessing optical flow presents unique spatiotemporal challenges. Static-image corruption suites already include conventional motion blur, but they do not explicitly model pair-specific effects such as asymmetric inter-frame exposure or video transmission artifacts. Prior optical-flow studies have examined adversarial attacks~\cite{ranjan2019attacking,agnihotri2024cospgd}, rendered corruption benchmarks such as RobustSpring~\cite{schmalfuss2025robustspring}, and broader reliability platforms such as FlowBench~\cite{agnihotri2025flowbench}. These studies motivate a complementary evaluation on captured source imagery with explicit inter-frame and video-pipeline corruptions.

To address this scope, we separate two complementary evaluation challenges: domain robustness under environmental and sensor-level degradations, and motion-distribution robustness under changes in inter-frame displacement. Based on this distinction, we introduce a real-video-based spatiotemporal corruption benchmark suite for optical flow (Fig.~\ref{fig:overview}), combining controlled corruptions, complementary stability and accuracy measurements, and two targeted benchmark datasets.

First, we assemble controlled corruptions tailored to optical-flow evaluation. In addition to established spatial degradations, seven temporal corruptions explicitly alter inter-frame consistency through asymmetric illumination, evolving weather, motion-induced blur, or video-pipeline failures. KITTI-FC and GoPro-FC retain objects, scene structure, and motion patterns from captured driving and egocentric data, while most corruption effects are algorithmically generated. This provides repeatable clean/corrupted pairs; it does not equate these effects with recorded adverse conditions.

To rigorously quantify the impact of these spatiotemporal corruptions, we use two complementary evaluation axes. The output-stability error $R^{s}$ measures the prediction shift between corrupted and clean inputs without requiring ground truth; it is closely aligned with RobustSpring's stability measure and is not claimed as a new standalone metric. The accuracy degradation $R^{d}$ uses ground truth to measure the signed change in EPE from clean to corrupted inputs. Their clean-error-normalized forms, $R^{s}_{\mathrm{rel}}$ and $R^{d}_{\mathrm{rel}}$, support proportional cross-model analysis when ground truth is available.

Finally, we systematically apply these corruptions to construct two distinct benchmark datasets, each targeting a specific robustness challenge. The first, KITTI-FC, is built upon KITTI-2015~\cite{menze2015object} to assess domain robustness from an ego-motion perspective. The second, GoPro-FC, is constructed from the high-frame-rate GoPro dataset~\cite{nah2017deep} to evaluate motion-distribution robustness. This dataset stress-tests a model's ability to handle the complex, erratic motions of dynamic agents and unforeseen events that fall outside the typical ego-motion distribution. KITTI-FC supports paired Out-of-Domain (OOD) and In-Domain (ID) evaluation, whereas GoPro-FC controls the tested motion distribution through 10/30 FPS sampling.

In summary, our key contributions are as follows:
\begin{itemize}
    \item We establish a real-video-based spatiotemporal corruption benchmark suite for optical-flow robustness that complements rendered and mixed-source benchmarks.
    \item We assemble 24 controlled corruptions, including seven temporal corruptions, for optical-flow evaluation.
    \item Our metrics combine GT-free output stability $R^{s}$, GT-based accuracy degradation $R^{d}$, clean-error-normalized ratios, and reference-normalized aggregation.
    \item We construct two complementary benchmark datasets, KITTI-FC and GoPro-FC, providing both domain-wise and motion-wise robustness evaluation capabilities.
    \item We conduct an extensive evaluation of 33 model variants, including the recent SEA-RAFT baseline, delivering evidence-bounded insights into the vulnerabilities of current models and providing actionable directions for developing robust optical flow estimators.
\end{itemize}

\section{Related Work}

\subsection{Optical Flow Estimation}
Optical flow estimation has experienced a paradigm shift with the advent of deep learning. While early approaches relied on hand-crafted energy minimization constraints~\cite{horn1981determining,farneback2003two,revaud2015epicflow}, modern supervised networks~\cite{sun2018pwc,ilg2017flownet,hui2018liteflownet} have substantially advanced accuracy. Iterative refinement architectures~\cite{teed2020raft,wang2025waft,morimitsu2025dpflow,jahedi2024ms,jahedi2024ccmr,sea_raft}, Transformer-based models with global context~\cite{huang2022flowformer,shi2023flowformer++}, and integrations with vision foundation models~\cite{zhou2024samflow} or diffusion processes~\cite{saxena2024surprising,luo2024flowdiffuser} achieve strong performance on standard benchmarks. Unsupervised methods~\cite{stone2021smurf,yuan2024unsamflow} address the scarcity of real-world ground-truth labels.

However, a limitation of this architectural evolution is its heavy reliance on idealized, clean evaluation benchmarks. Optical flow is a core perceptual component for safety-critical downstream applications such as visual odometry, robotic tracking, and dynamic scene understanding~\cite{qin2024rso,sormoli2024optical,liao2025sceneaware}. Despite this, the stability and reliability of these models under unconstrained degradations remain underexplored, motivating evaluation beyond clean-data accuracy.

\subsection{Optical-Flow Robustness and Adverse Conditions}
The vulnerability of deep neural networks to domain shifts and data corruptions is well documented. Corruption benchmarks are widely used for classification~\cite{hendrycks2019benchmarking,muller2023classification}, segmentation~\cite{kamann2020benchmarking,zhang2023delivering}, and object detection~\cite{michaelis2019benchmarking,zhu2023understanding}. Applying static-image protocols directly to optical flow, however, does not test failures that specifically disrupt inter-frame correspondence.

Optical-flow robustness has been studied from several complementary perspectives. Adversarial analyses probe worst-case perturbations~\cite{ranjan2019attacking,schmalfuss2022perturbation,koren2022consistent,schrodi2022towards}. Spring provides high-resolution rendered data with dense ground truth~\cite{mehl2023spring}, and RobustSpring extends it with time- and stereo-consistent corruptions and a clean-to-corrupted prediction-stability measure~\cite{schmalfuss2025robustspring}. FlowBench evaluates reliability and generalization across mixed source datasets, common corruptions, and adversarial attacks~\cite{agnihotri2025flowbench}. Flow-R instead modifies object shape and texture to analyze semantic matching sensitivity~\cite{long2025shape}. Other studies target adaptation to fog or multiple adverse-weather domains~\cite{zhou2023unsupervised,zhou2025adverse}, while ACDC provides directly captured adverse-condition driving scenes for semantic perception~\cite{sakaridis2021acdc}. These works address different aspects of reliability and motivate separating captured source content, generated corruption effects, and recorded adverse conditions.

Applying simulated corruptions to captured benchmark data is established in classification, segmentation, and autonomous-driving perception~\cite{hendrycks2019benchmarking,kamann2020benchmarking,dong2023benchmarking}. RobustSpring and FlowBench likewise apply controlled algorithmic corruptions to existing optical-flow benchmark data~\cite{schmalfuss2025robustspring,agnihotri2025flowbench}. As summarized in Tab.~\ref{tab:benchmark_comparison}, our joint protocol combines captured KITTI/GoPro content with asymmetric frame-pair and video-pipeline corruptions, motion control, paired ID/OOD evaluation, and GT-based degradation. These controls separate domain, temporal, motion, stability, and accuracy effects while retaining repeatable clean/corrupted pairs; they distinguish simulated corruptions from captured adverse conditions.

\begin{table*}[t]
\centering
\begingroup
\caption{\textbf{Comparison with closely related optical-flow robustness benchmarks.} A check mark indicates that the capability is explicitly included in the reported benchmark protocol.}
\label{tab:benchmark_comparison}
\scriptsize
\setlength{\tabcolsep}{4pt}
\renewcommand{\arraystretch}{1.15}
\begin{tabularx}{\textwidth}{@{}p{0.27\textwidth} >{\centering\arraybackslash\bfseries}X >{\centering\arraybackslash}X >{\centering\arraybackslash}X@{}}
\toprule
Aspect & Ours & RobustSpring & FlowBench \\
\midrule
Data source & Captured KITTI + GoPro & Rendered Spring & Mixed datasets \\
Asymmetric frame pairs & $\checkmark$ & $\checkmark$ & $\times$ \\
Video pipeline (H.264/bit error) & $\checkmark$ & $\times$ & $\times$ \\
Motion control & 10/30 FPS & $\times$ & $\times$ \\
Main severity reporting & Five-level average & One level & Level 3 \\
GT-free output stability & $R^{s}$: prediction shift & $R^{C}_{\mathrm{EPE}}$: prediction shift & $\times$ \\
GT-based clean-to-corrupted degradation & $R^{d}=\mathrm{EPE}_{\mathrm{corr}}-\mathrm{EPE}_{\mathrm{clean}}$ & $\times$ & $\times$ \\
Paired ID/OOD protocol & $\checkmark$ & $\times$ & $\times$ \\
\bottomrule
\end{tabularx}
\vskip -1.5\baselineskip plus -1fil
\endgroup
\end{table*}

\begin{figure*}
  \centering
  \includegraphics[width=\linewidth]{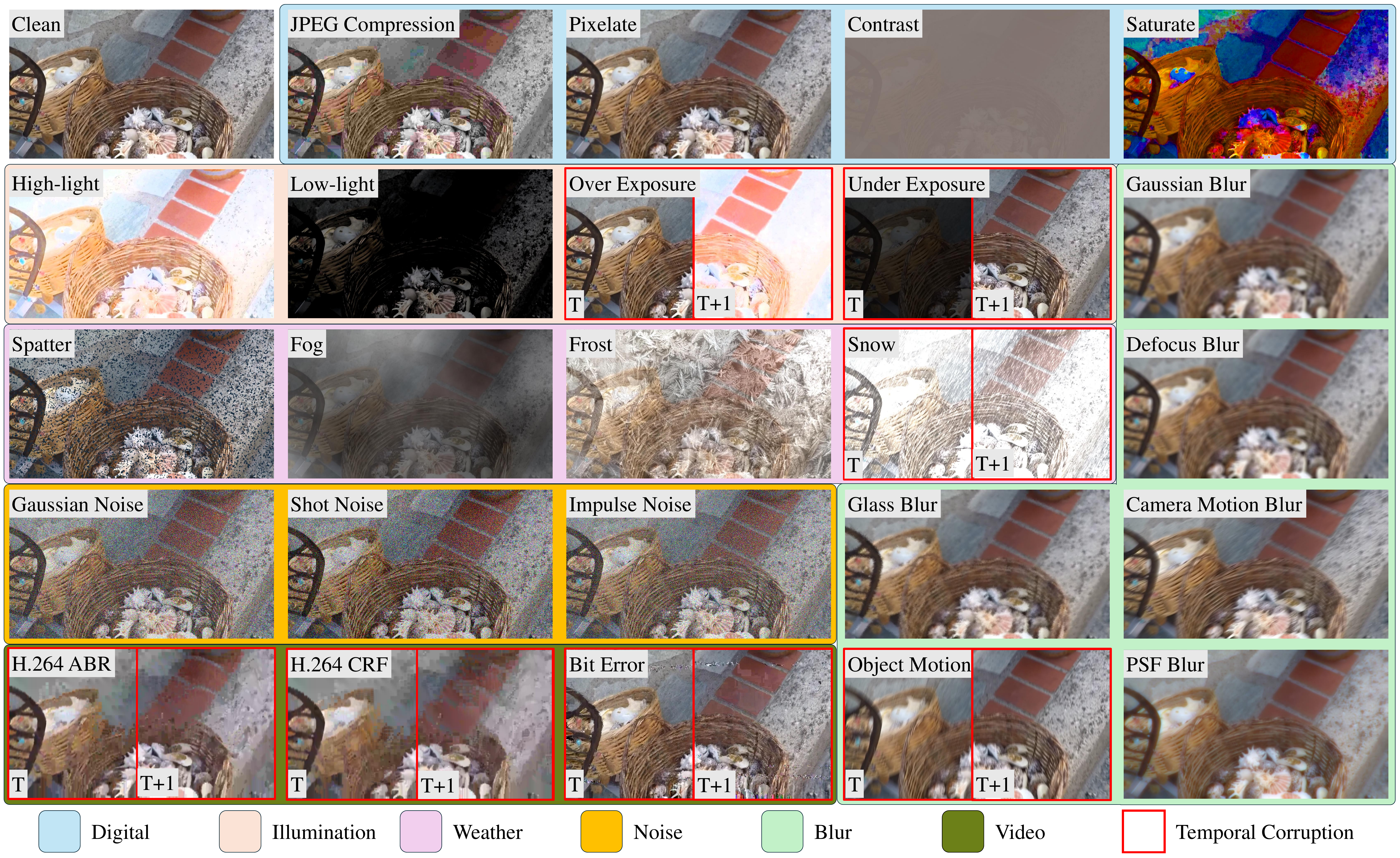}
  \caption{\textbf{Effects of the 24 corruptions at severity 5.}
  For temporal corruptions, we display the previous and next frames (left/right) to reveal inter-frame differences, which are crucial for diagnosing temporal robustness in optical flow. Examples are from GoPro-FC.}
  \label{fig:corruptions}
  \vskip -1.5\baselineskip plus -1fil
\end{figure*}

\section{Corruptions for Optical Flow}
\label{sec:corruptions}

A practical robustness benchmark for optical flow must cover both spatial degradation and disturbances that alter inter-frame correspondence. Our suite comprises $24$ corruption types in $6$ classes (Fig.~\ref{fig:corruptions}), each evaluated at $5$ severity levels.

Most corruption mechanisms and five-level schedules are adopted from established common-corruption benchmarks or published degradation protocols~\cite{hendrycks2019benchmarking,kamann2020benchmarking,dong2023benchmarking,michaelis2019benchmarking,yi2021benchmarking}. For temporal Over/Under Exposure and Snow, we retain the adopted per-frame degradation parameters but change the paired-frame rendering: exposure is applied asymmetrically and Snow evolves independently between frames. This converts established spatial degradations into inter-frame consistency tests without redefining their severity scale. Together with Object Motion Blur, H.264 CRF/ABR, and Bit Error, these form the seven temporal corruptions. Camera Motion Blur uses a common direction for both frames and is therefore categorized as blur rather than temporal corruption.

The four benchmark-selected schedules---Object Motion Blur, H.264 CRF, H.264 ABR, and Bit Error---are assessed on ten fixed pairs from five GoPro sequences. Because raw quality values are not directly comparable across mechanisms with different spatial support, we test within-mechanism progression over PSNR, SSIM, and normalized adjacent-frame-difference change. The selected schedules all have 12/12 expected-direction adjacent transitions and normalized trajectory deviations of 0.035--0.265. For comparison, the central 80\% of the $20$ adopted schedules spans 9--12/12 transitions and deviations of 0.057--0.299. Thus, the selected schedules follow the established ordinal trend distribution, without implying equal raw distortion across corruption mechanisms.

\subsection{Digital Corruptions}

Digital degradations arise from acquisition, compression, resampling, or distribution. They are pervasive in networked systems due to bandwidth constraints, storage limitations, and heterogeneous devices. Evaluating under these corruptions probes whether flow estimation remains reliable when the pipeline is optimized for throughput over fidelity.

\noindent \textbf{JPEG Compression.}
Each frame is independently compressed with JPEG using quality factors 25, 18, 15, 10, and 7 for severities 1 to 5. This corruption measures resilience to storage and transmission-oriented recompression that is common in fleet video logging and remote monitoring.

\noindent \textbf{Pixelate.}
We downsample with a box filter and rescale to the original resolution. The downsample factors are 0.60, 0.50, 0.40, 0.30, and 0.25. This simulates low-resolution streams from embedded cameras or adaptive streaming.

\noindent \textbf{Contrast.}
We adjust global contrast as
\begin{equation}
\hat{I}=\big(I-\operatorname{mean}(I)\big)\cdot c+\operatorname{mean}(I),
\end{equation}
with $c=0.40, 0.30, 0.20, 0.10, 0.05$. This tests stability when in-camera processing or post-processing alters scene contrast to meet display or compression budgets.

\noindent \textbf{Saturate.}
We convert to HSV and modify saturation by
\begin{equation}
\hat{S}=\mathrm{clip}(S\cdot \alpha + \beta),
\end{equation}
with $(\alpha,\beta)=(0.1,0), (0.3,0), (2,0), (5,0.1), (20,0.2)$. Values are clipped to the valid range. The goal is to ensure color shifts from ISP tuning or domain-specific color styles do not mislead motion estimation.

\subsection{Illumination Corruptions}

Illumination changes are ubiquitous in outdoor environments. Cameras traverse tunnels, underpasses, and shadows within fractions of a second. A robust system must handle both smooth global changes and exposure mismatches across frames when the camera cannot adapt quickly enough.

\noindent \textbf{High-Light.}
In HSV space, we increase brightness by an additive offset $c=0.10, 0.20, 0.30, 0.40, 0.50$ and clip to the valid range. This approximates global brightening caused by direct sunlight or reflective surfaces.

\noindent \textbf{Low-Light.}
We decrease brightness using the same offset schedule and clipping. This simulates dusk, tunnels, or poorly lit environments that frequently degrade motion cues.

\noindent \textbf{Over Exposure (temporal).}
One frame is over-exposed to emulate delayed metering during sudden brightening, using
\begin{equation}
\hat{V}= \mathrm{clip}\big(V\cdot 2^{\mathrm{ev}}\big),
\end{equation}
where $\mathrm{ev}=0.4, 0.8, 1.2, 1.6, 2.0$. This tests exposure bias that destabilizes matching and occlusion reasoning.

\noindent \textbf{Under Exposure (temporal).}
One frame is under-exposed using the same rule with $\mathrm{ev}=-0.4, -0.8, -1.2, -1.6, -2.0$. 
This case is critical for nighttime capture, where ISO and shutter trade-offs differ frame to frame.

\subsection{Weather Corruptions}

Weather is a primary risk factor for perception in outdoor environments. Rain, fog, frost, and snow obscure texture, alter contrast, and create transient occlusions. A robust flow estimator must preserve motion cues despite these effects.

\noindent \textbf{Spatter.}
We overlay adherent droplets with increasing occlusion ratios across severities. 
This models lens soiling and rain on housings, affecting roadside cameras and dashcams.

\noindent \textbf{Fog.}
We render an atmospheric veil and increase density with severity. This probes resilience to reduced visibility that weakens long-range motion and increases ambiguity.

\noindent \textbf{Frost.}
We overlay semi-transparent frost layers of increasing strength. This captures cold-start conditions where vehicles depart with partially frosted optics.

\noindent \textbf{Snow (temporal).}
We render moving particles and increase density and motion blur with severity. The two frames use different particle realizations while maintaining a consistent motion direction. This stresses matching under dynamic occluders that vary across frames, as encountered in snowfall.

\subsection{Noise Corruptions}

Electronic and photon noise increase substantially in low-light conditions or under fast exposure times. Robustness to noise ensures stable motion in safety-critical operations such as emergency braking and tracking.

\noindent \textbf{Gaussian Noise.}
\begin{equation}
\hat{I}= I + c\,n,\quad n\sim \mathcal{N}(0,1),
\end{equation}
with $c=0.08, 0.12, 0.18, 0.26, 0.38$. 
This corruption measures tolerance to sensor read noise and downstream amplification.

\noindent \textbf{Shot Noise.}
\begin{equation}
\hat{I}= \frac{\mathrm{Pois}(I\cdot c)}{c},
\end{equation}
with $c=60, 25, 12, 5, 3$. This tests robustness when photon count is low and variance scales with the signal, a common regime for low-light imaging.

\noindent \textbf{Impulse Noise.}
We replace a fraction of pixels by $0$ or $255$ with ratios $0.03$, $0.06$, $0.09$, $0.17$, and $0.27$. This simulates hot pixels and transient sensor glitches.

\begin{figure}
  \centering
  \includegraphics[width=\linewidth]{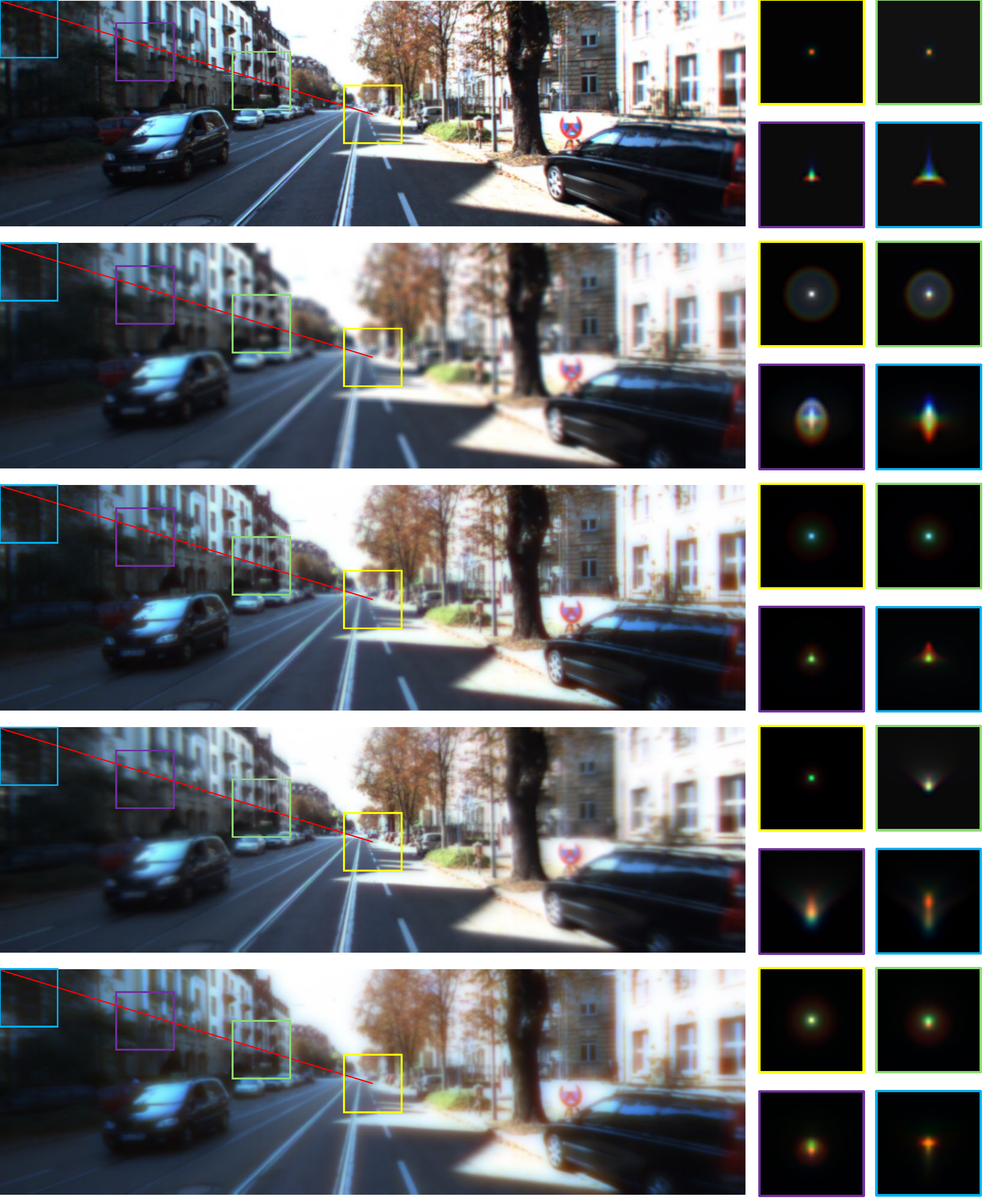}
  \caption{\textbf{PSFs at severity levels 1--5 (top to bottom).} Four field positions show spatial variation in blur strength and shape; enlarged PSF details appear on the right.}
  \label{fig:psf}
  \vskip -1.5\baselineskip plus -1fil
\end{figure}

\subsection{Blur Corruptions}

Blur arises from optics, focus, and motion. In real-world applications, it is caused by vibration, platform dynamics, and cost-constrained lenses. Evaluating with both analytic and physically grounded blur reveals whether an approach generalizes beyond simple kernels to real optical behavior.

\noindent \textbf{Gaussian Blur.}
We convolve with Gaussian kernels with standard deviations $1$, $2$, $3$, $4$, and $6$. 
This is a standard baseline for isotropic blur sensitivity.

\noindent \textbf{Defocus Blur.}
We convolve with a circular mean filter with radii $3$, $4$, $6$, $8$, and $10$ pixels. 
This represents misfocus due to rapid depth changes and focus sweep delays.

\noindent \textbf{Glass Blur.}
We first apply a Gaussian with standard deviation $\sigma$ and then shuffle pixels within a neighborhood of size $\alpha$ for $\beta$ iterations. We use $(\sigma,\alpha,\beta)=(0.7,1,2)$, $(0.9,2,1)$, $(1.0,2,3)$, $(1.1,3,2)$, and $(1.5,4,2)$. This tests robustness to locally permuted textures that confuse gradient-based matching.

\noindent \textbf{Camera Motion Blur.}
We sample a global motion direction that is shared by both frames and apply a Gaussian-weighted line spread with radius $\alpha$ and standard deviation $\sigma$. We use $(\alpha,\sigma)=(10,3)$, $(15,5)$, $(15,8)$, $(15,12)$, and $(20,15)$. This reflects platform shake that changes slowly in direction across adjacent frames in mobile platforms.

\noindent \textbf{Object Motion Blur (temporal).}
From native 240-FPS GoPro videos, we apply FLAVR~\cite{kalluri2023flavr} for 4$\times$ temporal interpolation. We then sample keyframes every 32 or 96 frames for the 30-FPS or 10-FPS setting, respectively, and average the $2c+1$ frames centered at each keyframe, where $c\in\{3,6,9,12,15\}$. This captures blur from fast-moving objects.

\noindent \textbf{PSF Blur.}
For each severity, we use spatially varying PSF fields from five real lenses designed by GSO~\cite{gao2024global}. PSF varies across the field of view, and we obtain the blurred image by convolving with the PSF field. The pixel width is 4~$\mu$m. The RMS spot radii of the lenses are 0.0296~mm, 0.0832~mm, 0.1102~mm, 0.1588~mm, and 0.1939~mm. This corruption evaluates sensitivity to real optical aberrations characteristic of cost-efficient consumer cameras, which cannot be fully captured by simple analytic kernels. An illustration of the corruption results are in Fig.~\ref{fig:psf}.

\subsection{Video Corruptions}

Video pipelines are indispensable in modern vision applications, from edge encoders on mobile devices to wireless backhaul and cloud archives. Compression and bitstream degradation directly challenge inter-frame consistency, so a robust flow estimator must tolerate these effects without catastrophic drift.

\noindent \textbf{H.264 CRF (temporal).}
We encode with Constant Rate Factor values 23, 30, 37, 44, and 51 for severities 1 to 5, following~\cite{yi2021benchmarking}. This simulates bandwidth-limited encoding.

\noindent \textbf{H.264 ABR (temporal).}
For interpolated GoPro-FC videos, we use average bitrates of 25, 12.5, 6.25, 3.125, and 1.5625 Mb/s. This setup evaluates robustness to rate control that trades fidelity for throughput in networked systems.

\noindent \textbf{Bit Error (temporal).}
We apply the FFmpeg bitstream filter with noise parameters 50M, 25M, 15M, 10M, and 5M on interpolated GoPro-FC videos. This creates controlled transmission-like decoding disruptions and is used as a diagnostic proxy rather than a direct recording of link failures.

\section{Metrics and Benchmarks}
\label{sec:benchmarks}

\subsection{Robustness Metrics}
\label{sec:metrics}

Real-world vision systems require optical flow estimators that remain dependable under visual and temporal degradations. While conventional End-Point Error (EPE) measures accuracy against ground truth, it does not separately reveal how a model's prediction changes when the input is corrupted.

Let $\mathbf{f}_{\mathrm{gt}}(p)\in\mathbb{R}^2$ be the ground-truth flow at valid pixel $p\in\Omega$, $\mathbf{f}_{\mathrm{clean}}(p)$ the prediction on a clean input pair, and $\mathbf{f}_{c,\ell}(p)$ the prediction under corruption type $c$ at severity level $\ell$. We define
\begin{equation}
\operatorname{EPE}(\mathbf{f}_a,\mathbf{f}_b)=
\frac{1}{|\Omega|}\sum_{p\in\Omega}\|\mathbf{f}_a(p)-\mathbf{f}_b(p)\|_2.
\end{equation}
The severity index is denoted by $\ell$ to distinguish it from the stability superscript in $R^{s}$.

\noindent\textbf{Output Stability Error.}
The GT-free output-stability error measures the prediction shift induced by corruption:
\begin{equation}
R^{s}_{c,\ell}=
\operatorname{EPE}\!\left(\mathbf{f}_{c,\ell},\mathbf{f}_{\mathrm{clean}}\right).
\label{eq:rs}
\end{equation}
A smaller $R^{s}$ indicates a more stable output. At the per-pixel EPE level, this is the same clean-to-corrupted prediction distance as RobustSpring's $R^{C}_{\mathrm{EPE}}$~\cite{schmalfuss2025robustspring}; accordingly, we do not claim $R^{s}$ as a new standalone metric. KITTI-FC evaluates it on the KITTI valid-flow mask, whereas GoPro-FC evaluates all pixels. It is useful without ground-truth optical flow.

\noindent\textbf{Accuracy Degradation.}
When ground truth is available, we measure the signed clean-to-corrupted change in accuracy:
\begin{equation}
R^{d}_{c,\ell}=
\operatorname{EPE}\!\left(\mathbf{f}_{c,\ell},\mathbf{f}_{\mathrm{gt}}\right)
-
\operatorname{EPE}\!\left(\mathbf{f}_{\mathrm{clean}},\mathbf{f}_{\mathrm{gt}}\right).
\label{eq:rd}
\end{equation}
Positive $R^{d}$ denotes accuracy degradation, whereas a negative value means that the corrupted input yields lower EPE than the clean input. Unlike corrupted-input EPE alone, $R^{d}$ explicitly isolates the change from each model's clean baseline.

\noindent\textbf{Clean-Error-Normalized Metrics.}
To express the two quantities relative to clean accuracy, we define
\begin{equation}
R^{s}_{\mathrm{rel}}=
\frac{R^{s}}{\operatorname{EPE}_{\mathrm{clean}}},
\qquad
R^{d}_{\mathrm{rel}}=
\frac{R^{d}}{\operatorname{EPE}_{\mathrm{clean}}}.
\label{eq:relative_metrics}
\end{equation}
Although $R^{s}$ itself is GT-free, $R^{s}_{\mathrm{rel}}$ requires ground truth because its denominator is the clean-input EPE. Smaller values indicate a smaller corruption effect relative to baseline error.

\begin{table}[t]
\centering
\begingroup
\caption{\textbf{Interpretation of the diagnostic metric suite.}}
\label{tab:metric_summary}
\scriptsize
\setlength{\tabcolsep}{2.5pt}
\renewcommand{\arraystretch}{1.12}
\begin{tabularx}{\columnwidth}{@{}l c c X@{}}
\toprule
Metric & GT & Clean-EPE norm. & Interpretation \\
\midrule
$R^{s}$ & No & No & Output shift; lower is more stable \\
$R^{d}$ & Yes & No & Signed accuracy degradation \\
$R^{s}_{\mathrm{rel}}$ & Yes & Yes & Relative output-stability error \\
$R^{d}_{\mathrm{rel}}$ & Yes & Yes & Relative accuracy degradation \\
\bottomrule
\end{tabularx}
\endgroup
\end{table}

\begin{figure}[t]
    \centering
    \includegraphics[width=\linewidth]{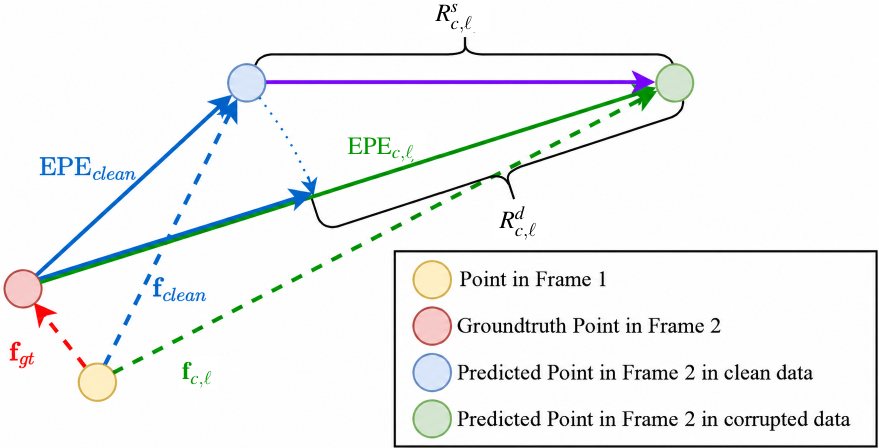}
    \caption{\textbf{Complementary robustness axes.} $R^{s}$ compares corrupted and clean predictions and does not require ground truth. $R^{d}$ uses flow GT to measure the signed EPE change from the clean baseline.}
    \label{fig:metric}
\end{figure}

\noindent\textbf{Reference-Normalized Aggregation.}
For model $m$ and corruption $c$, we first average $R^{s}$ over the five severity levels and then normalize by a fixed reference model:
\begin{equation}
\widetilde{R}^{s}_{m,c}=
\frac{\overline{R}^{s}_{m,c}}
{\overline{R}^{s}_{\mathrm{ref},c}}.
\label{eq:reference_normalization}
\end{equation}
We average $\widetilde{R}^{s}_{m,c}$ over corruptions to obtain a single ranking score. RAFT-OOD is used as the reference model because RAFT~\cite{teed2020raft} underpins many evaluated architectures and its OOD checkpoint represents a common deployment setting. The same reference-normalization procedure is applied to the other reported metric families where ground truth is available.
\subsection{Robustness Benchmarks}

Optical flow estimation is deployed across diverse visual sensing platforms—from mobile robots and autonomous agents to wearable cameras and surveillance systems. 
These environments are subject to degradations such as noise, blur, weather, exposure transitions, and video transmission artifacts that can destabilize temporal correspondence. 
To evaluate robustness across heterogeneous captured-source and controlled-degradation settings, we construct two complementary benchmarks: \textbf{KITTI-FC} for \emph{domain robustness} in dynamic outdoor scenes, and \textbf{GoPro-FC} for \emph{motion-distribution robustness} under complex egocentric dynamics.

\begin{figure}
\centering
\includegraphics[width=\linewidth]{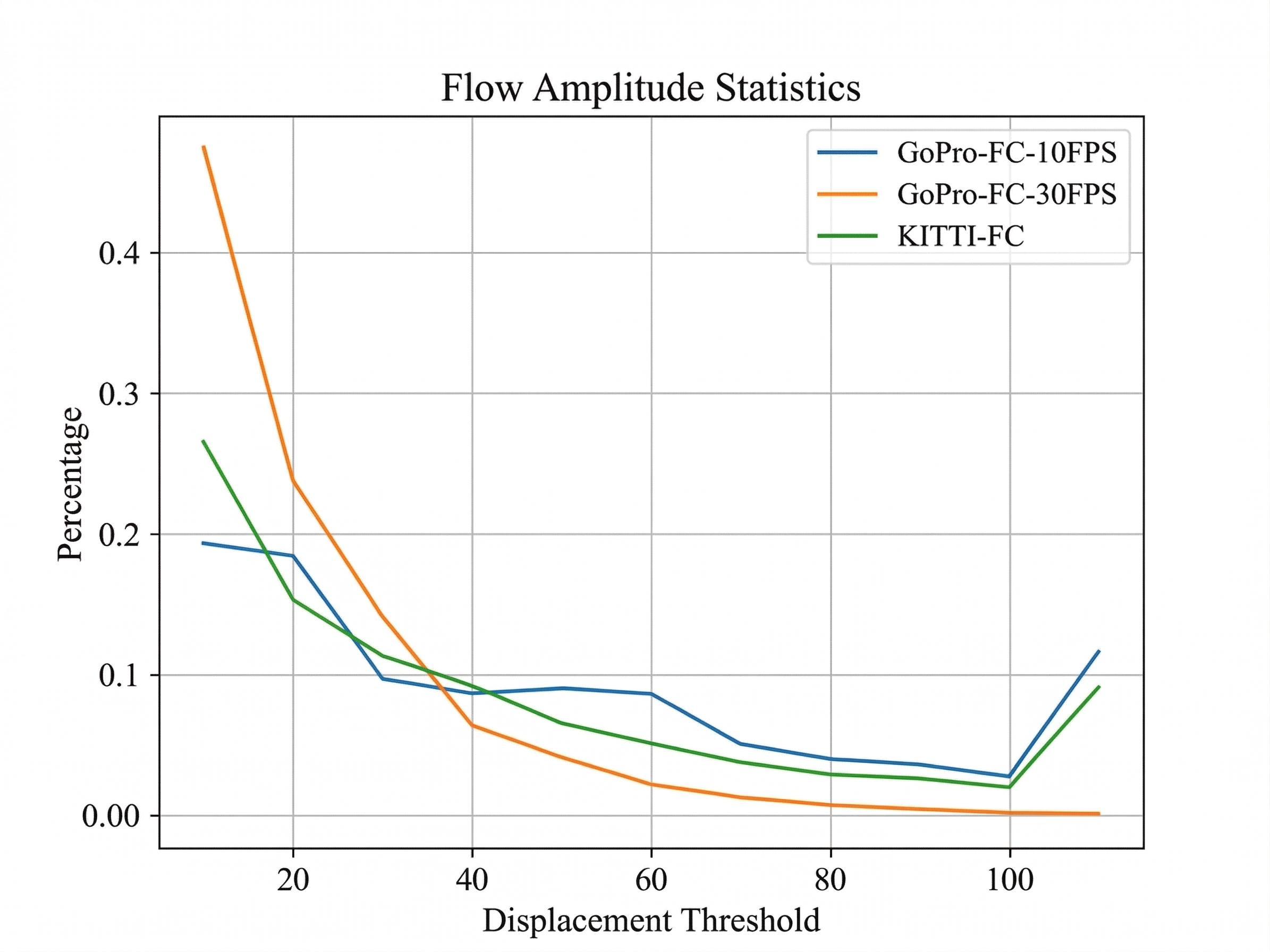}
\caption{\textbf{Optical flow amplitude statistics of KITTI-FC and GoPro-FC.} 
GoPro-FC-30FPS exhibits more small displacements, while GoPro-FC-10FPS has displacements similar to KITTI-FC.}
\label{fig:kitti_gopro_statistic}
\end{figure}

\noindent\textbf{KITTI-FC.}
The KITTI-FC benchmark is designed to assess the robustness of optical flow models in outdoor navigation scenarios. 
It is constructed based on the KITTI-2015 optical flow dataset~\cite{menze2015object}, which contains street-view image pairs captured at 10\,Hz. 
To investigate both \textit{in-domain} (ID) and \textit{out-of-domain} (OOD) robustness, the 200 training image pairs are split into 120 pairs for ID fine-tuning and 80 pairs for testing.
During corruption, we apply 20 corruption types proposed in Sec.~\ref{sec:corruptions} to the testing subset. 
The \textit{Object Motion Blur} and three video-related corruptions are excluded, as the limited frame count of KITTI prevents realistic temporal synthesis and may introduce artifacts. 
KITTI-FC thus provides a controlled outdoor setting to assess how models handle environmental and sensor-level degradations typical of mobile platform cameras.

Furthermore, since prior researches~\cite{ma2024posebench,yan2024benchmarking} have demonstrated that the inclusion of data augmentation during training bolsters model robustness, the KITTI-FC training set furnishes these corruptions, thereby providing a robust data foundation for the training of optical flow models. These corrupted training images are provided for potential robustness training; the supervised ID fine-tuning reported here uses only the clean versions of the 120 training pairs, with standard training augmentations.

\noindent\textbf{GoPro-FC.}
To evaluate robustness under diverse motion patterns and temporal disturbances characteristic of egocentric sensing in complex environments (\textit{e.g}., human motion, mobile robots, and other dynamic agents), we construct GoPro-FC from the high-frame-rate GoPro dataset~\cite{nah2017deep}.  
We select five representative video sequences (\textit{GOPR0374\_11\_02}, \textit{GOPR0379\_11\_00}, \textit{GOPR0384\_11\_02}, \textit{GOPR0385\_11\_00}, and \textit{GOPR0386\_11\_00}) to construct the benchmark. 
To simulate different motion magnitudes, frames are sampled at two temporal rates: 30\,FPS and 10\,FPS, forming two versions, \textit{GoPro-FC-30FPS} and \textit{GoPro-FC-10FPS}, respectively. 
Each sequence yields 135 image pairs for the 30\,FPS version and 43 pairs for the 10\,FPS version. 
To support video-based corruptions, we first convert raw GoPro sequences into videos via \texttt{FFMPEG}, perform $4\times$ temporal interpolation using FLAVR~\cite{kalluri2023flavr}, and then apply the 24 corruptions defined in Sec.~\ref{sec:corruptions}. 
Among these, video corruptions are directly applied on the interpolated 960\,FPS videos, while non-video corruptions operate on clean image pairs. 
As the GoPro dataset does not provide optical-flow ground truth, GoPro-FC supports only the GT-free output-stability error $R^{s}$; $R^{d}$ and both clean-error-normalized metrics require ground truth and are omitted.

\noindent\textbf{Comparison and Analysis.}
Figure~\ref{fig:kitti_gopro_statistic} quantifies the motion differences using RAFT-estimated flow magnitudes~\cite{shi2024beyond}: KITTI-FC at 10\,Hz typically has larger displacements than GoPro-FC-30FPS, whereas GoPro-FC-10FPS approximates KITTI-FC.

\noindent\textbf{Cross-benchmark comparison at similar SSIM.} RobustSpring sets one severity per corruption by balancing frame-wise SSIM on rendered RobustSpring~\cite{schmalfuss2025robustspring}; SSIM does not match displacement, texture, motion-boundary, or occlusion statistics across datasets. Over all 80 KITTI-FC validation pairs, JPEG-L5, Spatter-L4, and Glass-Blur-L1 yield mean SSIM $0.694/0.731/0.729$, close to $0.70/0.72/0.73$ in RobustSpring, yet under the original checkpoint settings, all six RAFT/GMA prediction shifts are $2.0$--$6.4\times$ larger on KITTI-FC. This comparison is descriptive: similar SSIM does not control for differences in datasets, model checkpoints, or corruption implementations, so the larger prediction shifts cannot be attributed solely to captured-scene content.

\section{Benchmark Results}

\begin{table}
\centering
\caption{\textbf{Evaluated methods.} Encoder and decoder architectures are listed where applicable.}
\resizebox{0.5\textwidth}{!}
{
\begin{tabular}{lcccc}
\Xhline{3\arrayrulewidth}
\textbf{Model} & \textbf{Type} & \textbf{Encoder} & \textbf{Decoder} \\ \Xhline{3\arrayrulewidth}
\textit{Farneb{\"a}ck}~\cite{farneback2003two} & Knowledge-driven & - & -  \\
EpicFlow~\cite{revaud2015epicflow} & Knowledge-driven & - & -  \\ 
DIS~\cite{kroeger2016fast} & Knowledge-driven & - & -  \\ \hline
RAFT~\cite{teed2020raft} & Supervised & CNN & GRU  \\
GMA~\cite{jiang2021learning} & Supervised & CNN & GRU  \\
CSFlow~\cite{shi2022csflow} & Supervised & CNN & GRU  \\
SKFlow~\cite{sun2022skflow} & Supervised & CNN & GRU  \\
GMFlowNet~\cite{zhao2022global} & Supervised & CNN & GRU  \\
CRAFT~\cite{sui2022craft} & Supervised & CNN\&Transformer & GRU  \\
FlowFormer~\cite{huang2022flowformer} & Supervised & Transformer & GRU  \\
FlowFormer++~\cite{shi2023flowformer++} & Supervised & Transformer & GRU  \\
SEA-RAFT~\cite{sea_raft} & Supervised & CNN & Recurrent  \\
SAMFlow~\cite{zhou2024samflow} & Supervised & Transformer\&SAM & GRU  \\
FlowDiffuser~\cite{luo2024flowdiffuser} & Supervised & Transformer & Diffusion  \\
DPFlow~\cite{morimitsu2025dpflow} & Supervised & CNN & GRU  \\\hline

SMURF~\cite{stone2021smurf} & Unsupervised & CNN & GRU  \\
BrightFlow~\cite{marsal2023brightflow} & Unsupervised & CNN & GRU  \\ \Xhline{3\arrayrulewidth}
\end{tabular}
}

\vskip -1.5\baselineskip plus -1fil
\label{tab:models}
\end{table}

\subsection{Implementation Details}
We evaluate $33$ model variants from $17$ representative methods, as categorized in Tab.~\ref{tab:models}.

\subsubsection{Knowledge-driven methods}
\noindent \textbf{Gunnar Farnebäck's algorithm:}
We use the OpenCV implementation with the default parameters for Gunnar Farnebäck's algorithm~\cite{farneback2003two}.
\noindent \textbf{EpicFlow:}
Following the official implementation, we run EpicFlow~\cite{revaud2015epicflow} by extracting edges and matches with SED~\cite{dollar2013structured} and DeepMatching~\cite{weinzaepfel2013deepflow}, respectively, and then estimating flow with the released EpicFlow code.
\noindent \textbf{DIS:}
The DIS~\cite{kroeger2016fast} is implemented with its RGB version following the official code.

\begin{figure*}
    \centering
    \includegraphics[width=0.99\linewidth]{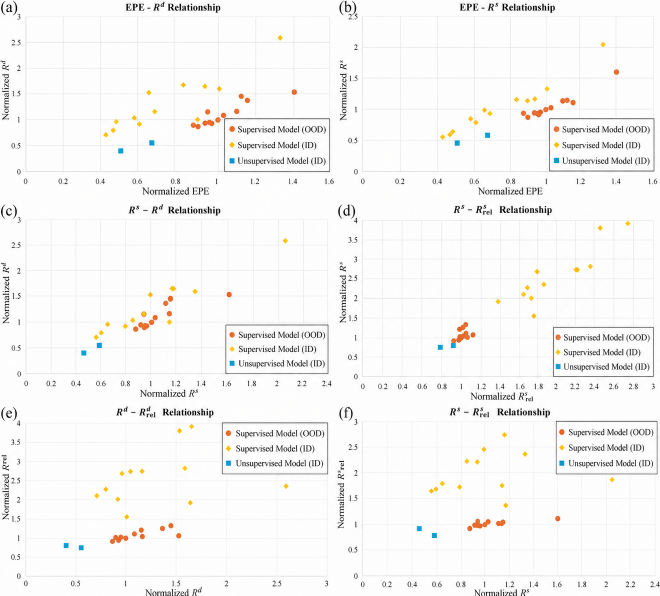}
    \caption{\textbf{Relationships among clean EPE, output stability, and accuracy degradation.} The plots compare EPE, $R^{s}$, $R^{d}$, $R^{s}_{\mathrm{rel}}$, and $R^{d}_{\mathrm{rel}}$ after reference normalization. The observed associations are descriptive and do not establish architectural causality.}
    \label{fig:metric_chart}
    \vskip -1.5\baselineskip plus -1fil
\end{figure*}

\begin{table*}[!t]
\centering
\caption{\textbf{$R^{s}$ results on KITTI-FC.} HL: \textit{High-light}, LL: \textit{Low-light}, OE: \textit{Over Exposure}, UE: \textit{Under Exposure}.}
\resizebox{\linewidth}{!}{

\begin{tabular}{lcccccccccccccccccccccc}
\Xhline{3\arrayrulewidth}
\multicolumn{1}{l|}{\multirow{2}{*}{\raisebox{-4ex}{\textbf{Model}}}} & \multicolumn{1}{c|}{\multirow{2}{*}{\raisebox{-4ex}{\textbf{Clean}}}} & \multicolumn{4}{c|}{\textbf{Digital}} & \multicolumn{4}{c|}{\textbf{Illumination}} & \multicolumn{4}{c|}{\textbf{Weather}} & \multicolumn{3}{c|}{\textbf{Noise}} & \multicolumn{5}{c|}{\textbf{Blur}} & \multicolumn{1}{c}{\multirow{2}{*}{\rotatebox{90}{RANK~~~~}}} \\ \cline{3-22} 
\multicolumn{1}{l|}{} & \multicolumn{1}{c|}{} & \rotatebox{90}{JPEG} & \rotatebox{90}{Pixelate} & \rotatebox{90}{Contrast} & \multicolumn{1}{c|}{\rotatebox{90}{Saturate}} & \rotatebox{90}{HL\tnote{*}} & \rotatebox{90}{LL\tnote{*}} & \rotatebox{90}{OE\tnote{*}} & \multicolumn{1}{c|}{\rotatebox{90}{UE\tnote{*}}} & \rotatebox{90}{Spatter} & \rotatebox{90}{Fog} & \rotatebox{90}{Frost} & \multicolumn{1}{c|}{\rotatebox{90}{Snow}} & \rotatebox{90}{Gaussian} & \rotatebox{90}{Shot} & \multicolumn{1}{c|}{\rotatebox{90}{Impulse}} & \rotatebox{90}{Gaussian} & \rotatebox{90}{Defocus} & \rotatebox{90}{Glass} & \rotatebox{90}{Camera} & \multicolumn{1}{c|}{\rotatebox{90}{PSF}}\\ \Xhline{3\arrayrulewidth}
\multicolumn{22}{c}{\textit{Knowledge-driven models}} \\ \Xhline{3\arrayrulewidth}
\multicolumn{1}{l|}{\textit{Farneb{\"a}ck}} & \multicolumn{1}{c|}{25.6} & 2.3 & 2.4 & 10.2 & \multicolumn{1}{c|}{3.3} & 4.0 & 4.5 & 7.5 & \multicolumn{1}{c|}{6.0} & 6.2 & 8.0 & 8.1 & \multicolumn{1}{c|}{9.7} & 5.5 & 5.5 & \multicolumn{1}{c|}{6.0} & 8.1 & 8.7 & 5.8 & 6.6 & \multicolumn{1}{c|}{7.8} & 3 \\
\multicolumn{1}{l|}{EpicFlow} & \multicolumn{1}{c|}{9.4} & 7.4 & 2.3 & 5.4 & \multicolumn{1}{c|}{4.0} & 3.2 & 6.4 & 8.4 & \multicolumn{1}{c|}{5.6} & 16.8 & 5.0 & 21.6 & \multicolumn{1}{c|}{13.0} & 20.1 & 26.2 & \multicolumn{1}{c|}{18.5} & 3.2 & 3.5 & 5.2 & 2.8 & \multicolumn{1}{c|}{3.1} & 2 \\
\multicolumn{1}{l|}{DIS} & \multicolumn{1}{c|}{20.6} & 1.9 & 0.9 & 1.9 & \multicolumn{1}{c|}{2.3} & 2.7 & 3.7 & 8.0 & \multicolumn{1}{c|}{10.8} & 5.4 & 5.8 & 8.9 & \multicolumn{1}{c|}{6.5} & 2.5 & 2.3 & \multicolumn{1}{c|}{2.7} & 1.6 & 1.8 & 2.0 & 2.3 & \multicolumn{1}{c|}{2.0} & 1 \\\Xhline{3\arrayrulewidth}
\multicolumn{22}{c}{\textit{Out-Of-Domain (OOD) Supervised Models}} \\ \Xhline{3\arrayrulewidth}
\multicolumn{1}{l|}{RAFT} & \multicolumn{1}{c|}{4.3} & 7.6 & 1.5 & 1.5 & \multicolumn{1}{c|}{2.4} & 2.0 & 4.2 & 3.3 & \multicolumn{1}{c|}{0.4} & 18.6 & 3.2 & 25.2 & \multicolumn{1}{c|}{12.2} & 11.0 & 9.3 & \multicolumn{1}{c|}{11.5} & 3.4 & 3.8 & 6.4 & 2.8 & \multicolumn{1}{c|}{2.7} & 7 \\
\multicolumn{1}{l|}{GMA} & \multicolumn{1}{c|}{4.2} & 7.5 & 1.5 & 1.3 & \multicolumn{1}{c|}{2.5} & 1.8 & 4.5 & 2.9 & \multicolumn{1}{c|}{0.4} & 20.8 & 2.9 & 26.6 & \multicolumn{1}{c|}{12.0} & 12.6 & 10.8 & \multicolumn{1}{c|}{13.8} & 3.6 & 4.2 & 7.6 & 2.5 & \multicolumn{1}{c|}{2.7} & 8 \\
\multicolumn{1}{l|}{CSFlow} & \multicolumn{1}{c|}{4.1} & 6.9 & 1.2 & 1.7 & \multicolumn{1}{c|}{2.0} & 2.0 & 4.8 & 3.3 & \multicolumn{1}{c|}{0.5} & 19.0 & 3.2 & 25.4 & \multicolumn{1}{c|}{11.9} & 8.6 & 7.2 & \multicolumn{1}{c|}{9.0} & 3.1 & 3.6 & 6.1 & 2.8 & \multicolumn{1}{c|}{2.5} & 5 \\
\multicolumn{1}{l|}{SKFlow} & \multicolumn{1}{c|}{4.0} & 6.3 & 1.0 & 1.3 & \multicolumn{1}{c|}{2.2} & 1.5 & 3.8 & 2.6 & \multicolumn{1}{c|}{0.3} & 18.1 & 2.4 & 24.1 & \multicolumn{1}{c|}{11.1} & 12.1 & 10.8 & \multicolumn{1}{c|}{13.1} & 3.3 & 3.8 & 6.4 & 2.4 & \multicolumn{1}{c|}{2.7} & 2 \\
\multicolumn{1}{l|}{GMFlowNet} & \multicolumn{1}{c|}{4.1} & 6.7 & 1.4 & 1.6 & \multicolumn{1}{c|}{2.1} & 1.9 & 5.0 & 2.8 & \multicolumn{1}{c|}{0.6} & 18.7 & 3.1 & 24.5 & \multicolumn{1}{c|}{9.1} & 9.9 & 8.6 & \multicolumn{1}{c|}{10.5} & 2.9 & 3.3 & 5.5 & 2.4 & \multicolumn{1}{c|}{2.4} & 4 \\
\multicolumn{1}{l|}{CRAFT} & \multicolumn{1}{c|}{4.7} & 7.2 & 1.4 & 1.8 & \multicolumn{1}{c|}{2.9} & 2.0 & 4.5 & 3.3 & \multicolumn{1}{c|}{0.5} & 19.8 & 3.2 & 26.9 & \multicolumn{1}{c|}{15.3} & 13.8 & 12.3 & \multicolumn{1}{c|}{14.6} & 3.6 & 4.0 & 8.7 & 2.8 & \multicolumn{1}{c|}{2.8} & 11 \\
\multicolumn{1}{l|}{FlowFormer} & \multicolumn{1}{c|}{4.8} & 6.5 & 1.6 & 3.5 & \multicolumn{1}{c|}{2.9} & 2.1 & 5.0 & 3.2 & \multicolumn{1}{c|}{1.0} & 19.8 & 3.4 & 25.5 & \multicolumn{1}{c|}{12.0} & 9.7 & 8.4 & \multicolumn{1}{c|}{10.6} & 3.6 & 3.7 & 7.0 & 2.7 & \multicolumn{1}{c|}{2.7} & 12 \\
\multicolumn{1}{l|}{FlowFormer++} & \multicolumn{1}{c|}{4.7} & 7.4 & 2.3 & 2.9 & \multicolumn{1}{c|}{3.0} & 2.3 & 5.4 & 2.8 & \multicolumn{1}{c|}{0.8} & 20.0 & 3.2 & 26.2 & \multicolumn{1}{c|}{11.2} & 10.6 & 9.7 & \multicolumn{1}{c|}{11.7} & 3.5 & 3.6 & 8.0 & 2.7 & \multicolumn{1}{c|}{2.6} & 13 \\
\multicolumn{1}{l|}{SAMFlow-T} & \multicolumn{1}{c|}{4.2} & 6.7 & 1.5 & 2.7 & \multicolumn{1}{c|}{2.7} & 1.8 & 3.7 & 2.6 & \multicolumn{1}{c|}{0.4} & 18.6 & 2.4 & 24.9 & \multicolumn{1}{c|}{10.1} & 10.2 & 8.9 & \multicolumn{1}{c|}{9.8} & 3.1 & 3.3 & 6.1 & 2.6 & \multicolumn{1}{c|}{2.5} & 6 \\
\multicolumn{1}{l|}{SAMFlow-B} & \multicolumn{1}{c|}{4.1} & 5.9 & 1.5 & 2.2 & \multicolumn{1}{c|}{2.3} & 1.7 & 3.6 & 2.4 & \multicolumn{1}{c|}{0.4} & 18.7 & 2.4 & 24.8 & \multicolumn{1}{c|}{8.7} & 8.1 & 7.3 & \multicolumn{1}{c|}{7.9} & 3.0 & 3.3 & 6.2 & 2.2 & \multicolumn{1}{c|}{2.2} & 1 \\
\multicolumn{1}{l|}{SAMFlow-H} & \multicolumn{1}{c|}{3.8} & 5.2 & 1.8 & 2.4 & \multicolumn{1}{c|}{2.7} & 2.0 & 3.8 & 2.7 & \multicolumn{1}{c|}{0.4} & 18.6 & 2.5 & 24.9 & \multicolumn{1}{c|}{8.6} & 9.5 & 8.4 & \multicolumn{1}{c|}{8.5} & 3.0 & 3.3 & 6.1 & 2.3 & \multicolumn{1}{c|}{2.5} & 3 \\
\multicolumn{1}{l|}{FlowDiffuser} & \multicolumn{1}{c|}{6.2} & 10.4 & 2.8 & 3.7 & \multicolumn{1}{c|}{4.0} & 4.2 & 5.7 & 4.6 & \multicolumn{1}{c|}{1.5} & 18.4 & 4.0 & 25.2 & \multicolumn{1}{c|}{18.8} & 15.6 & 13.6 & \multicolumn{1}{c|}{16.3} & 5.4 & 5.4 & 7.6 & 4.4 & \multicolumn{1}{c|}{4.1} & 14 \\
\multicolumn{1}{l|}{DPFlow} & \multicolumn{1}{c|}{3.8} & 6.4 & 1.5 & 2.4 & \multicolumn{1}{c|}{1.8} & 1.6 & 3.8 & 2.4 & \multicolumn{1}{c|}{0.6} & 18.2 & 4.1 & 24.3 & \multicolumn{1}{c|}{11.9} & 6.5 & 4.9 & \multicolumn{1}{c|}{7.1} & 6.7 & 5.4 & 7.4 & 2.9 & \multicolumn{1}{c|}{4.1} & 9 \\
\multicolumn{1}{l|}{SEA-RAFT} & \multicolumn{1}{c|}{3.8} & 8.4 & 1.4 & 1.9 & \multicolumn{1}{c|}{1.9} & 1.9 & 3.6 & 3.1 & \multicolumn{1}{c|}{0.9} & 18.4 & 3.7 & 24.4 & \multicolumn{1}{c|}{14.0} & 8.5 & 7.1 & \multicolumn{1}{c|}{9.4} & 4.1 & 4.5 & 8.3 & 2.6 & \multicolumn{1}{c|}{3.2} & 10 \\ \Xhline{3\arrayrulewidth}
\multicolumn{22}{c}{\textit{In-Domain (ID) Supervised Models}} \\ \Xhline{3\arrayrulewidth}
\multicolumn{1}{l|}{RAFT} & \multicolumn{1}{c|}{1.8} & 4.6 & 0.8 & 3.2 & \multicolumn{1}{c|}{1.4} & 2.3 & 5.8 & 6.5 & \multicolumn{1}{c|}{0.3} & 11.3 & 2.5 & 20.0 & \multicolumn{1}{c|}{35.2} & 15.2 & 10.1 & \multicolumn{1}{c|}{14.1} & 3.6 & 3.5 & 3.6 & 4.5 & \multicolumn{1}{c|}{3.3} & 11 \\
\multicolumn{1}{l|}{GMA} & \multicolumn{1}{c|}{2.4} & 4.2 & 1.5 & 2.5 & \multicolumn{1}{c|}{1.5} & 4.6 & 3.4 & 6.3 & \multicolumn{1}{c|}{0.4} & 14.2 & 3.0 & 22.4 & \multicolumn{1}{c|}{51.1} & 21.1 & 13.8 & \multicolumn{1}{c|}{20.0} & 3.1 & 3.3 & 3.9 & 4.6 & \multicolumn{1}{c|}{2.6} & 13 \\
\multicolumn{1}{l|}{CSFlow} & \multicolumn{1}{c|}{2.0} & 2.9 & 0.9 & 1.9 & \multicolumn{1}{c|}{1.4} & 1.8 & 4.0 & 2.8 & \multicolumn{1}{c|}{0.3} & 10.0 & 2.3 & 18.3 & \multicolumn{1}{c|}{16.8} & 10.7 & 7.7 & \multicolumn{1}{c|}{12.4} & 2.4 & 2.5 & 2.5 & 2.4 & \multicolumn{1}{c|}{2.0} & 5 \\
\multicolumn{1}{l|}{SKFlow} & \multicolumn{1}{c|}{2.8} & 3.5 & 1.2 & 1.8 & \multicolumn{1}{c|}{1.2} & 3.2 & 3.3 & 5.0 & \multicolumn{1}{c|}{0.3} & 12.7 & 2.3 & 20.4 & \multicolumn{1}{c|}{38.4} & 22.1 & 15.3 & \multicolumn{1}{c|}{23.8} & 2.8 & 3.1 & 3.5 & 3.0 & \multicolumn{1}{c|}{2.4} & 10 \\
\multicolumn{1}{l|}{GMFlowNet} & \multicolumn{1}{c|}{1.8} & 2.8 & 0.8 & 2.0 & \multicolumn{1}{c|}{1.0} & 1.6 & 3.4 & 7.6 & \multicolumn{1}{c|}{0.4} & 11.7 & 2.4 & 19.8 & \multicolumn{1}{c|}{21.8} & 14.1 & 11.7 & \multicolumn{1}{c|}{15.4} & 2.6 & 2.7 & 2.8 & 2.3 & \multicolumn{1}{c|}{2.3} & 8 \\
\multicolumn{1}{l|}{CRAFT} & \multicolumn{1}{c|}{3.7} & 4.7 & 1.7 & 3.4 & \multicolumn{1}{c|}{2.2} & 2.4 & 4.8 & 2.9 & \multicolumn{1}{c|}{0.8} & 14.5 & 4.0 & 24.0 & \multicolumn{1}{c|}{32.0} & 13.0 & 9.8 & \multicolumn{1}{c|}{12.8} & 3.3 & 3.5 & 4.8 & 3.0 & \multicolumn{1}{c|}{2.5} & 12 \\
\multicolumn{1}{l|}{FlowFormer} & \multicolumn{1}{c|}{1.7} & 2.8 & 1.3 & 3.7 & \multicolumn{1}{c|}{1.2} & 1.7 & 3.6 & 2.9 & \multicolumn{1}{c|}{0.8} & 15.7 & 3.1 & 22.1 & \multicolumn{1}{c|}{21.4} & 8.0 & 5.9 & \multicolumn{1}{c|}{8.8} & 3.6 & 3.6 & 3.3 & 2.7 & \multicolumn{1}{c|}{2.8} & 9 \\
\multicolumn{1}{l|}{FlowFormer++} & \multicolumn{1}{c|}{1.6} & 3.1 & 1.6 & 2.6 & \multicolumn{1}{c|}{1.3} & 1.3 & 3.4 & 2.1 & \multicolumn{1}{c|}{0.6} & 16.3 & 2.3 & 23.1 & \multicolumn{1}{c|}{12.1} & 9.0 & 6.9 & \multicolumn{1}{c|}{8.5} & 3.0 & 3.2 & 3.6 & 2.3 & \multicolumn{1}{c|}{2.5} & 7 \\
\multicolumn{1}{l|}{SAMFlow-T} & \multicolumn{1}{c|}{1.5} & 2.4 & 1.0 & 2.9 & \multicolumn{1}{c|}{0.9} & 1.0 & 2.4 & 1.2 & \multicolumn{1}{c|}{0.4} & 16.2 & 1.9 & 21.6 & \multicolumn{1}{c|}{10.6} & 4.4 & 3.7 & \multicolumn{1}{c|}{4.6} & 2.1 & 2.2 & 2.7 & 2.0 & \multicolumn{1}{c|}{1.8} & 4 \\
\multicolumn{1}{l|}{SAMFlow-B} & \multicolumn{1}{c|}{1.5} & 2.6 & 1.0 & 2.2 & \multicolumn{1}{c|}{0.8} & 1.3 & 2.4 & 1.1 & \multicolumn{1}{c|}{0.2} & 16.8 & 1.7 & 22.0 & \multicolumn{1}{c|}{9.1} & 5.1 & 3.8 & \multicolumn{1}{c|}{4.7} & 1.7 & 1.9 & 3.2 & 1.9 & \multicolumn{1}{c|}{1.5} & 3 \\
\multicolumn{1}{l|}{SAMFlow-H} & \multicolumn{1}{c|}{1.5} & 2.1 & 1.1 & 1.8 & \multicolumn{1}{c|}{0.8} & 1.0 & 2.2 & 1.1 & \multicolumn{1}{c|}{0.3} & 16.2 & 1.6 & 21.3 & \multicolumn{1}{c|}{10.0} & 4.0 & 2.6 & \multicolumn{1}{c|}{4.0} & 1.7 & 1.9 & 2.9 & 1.7 & \multicolumn{1}{c|}{1.4} & 2 \\
\multicolumn{1}{l|}{FlowDiffuser} & \multicolumn{1}{c|}{4.7} & 6.5 & 2.6 & 5.8 & \multicolumn{1}{c|}{4.5} & 4.7 & 8.8 & 7.7 & \multicolumn{1}{c|}{3.0} & 15.6 & 4.7 & 23.6 & \multicolumn{1}{c|}{31.0} & 26.2 & 22.6 & \multicolumn{1}{c|}{26.3} & 4.3 & 4.5 & 6.0 & 3.8 & \multicolumn{1}{c|}{3.9} & 14 \\
\multicolumn{1}{l|}{DPFlow} & \multicolumn{1}{c|}{1.5} & 1.9 & 0.5 & 0.9 & \multicolumn{1}{c|}{0.5} & 0.6 & 3.1 & 0.8 & \multicolumn{1}{c|}{0.1} & 11.8 & 1.8 & 17.7 & \multicolumn{1}{c|}{11.1} & 3.3 & 1.5 & \multicolumn{1}{c|}{2.8} & 1.8 & 1.8 & 2.1 & 1.7 & \multicolumn{1}{c|}{1.6} & 1 \\
\multicolumn{1}{l|}{SEA-RAFT} & \multicolumn{1}{c|}{1.7} & 3.9 & 0.7 & 1.5 & \multicolumn{1}{c|}{0.7} & 1.3 & 2.9 & 2.7 & \multicolumn{1}{c|}{0.5} & 9.9 & 1.6 & 17.2 & \multicolumn{1}{c|}{43.6} & 3.2 & 2.4 & \multicolumn{1}{c|}{3.3} & 2.1 & 2.2 & 2.5 & 5.5 & \multicolumn{1}{c|}{1.9} & 6 \\\Xhline{3\arrayrulewidth}
\multicolumn{22}{c}{\textit{In-Domain (ID) Unsupervised Models}} \\ \Xhline{3\arrayrulewidth}
\multicolumn{1}{l|}{SMURF} & \multicolumn{1}{c|}{2.1} & 1.9 & 0.5 & 0.3 & \multicolumn{1}{c|}{0.6} & 0.9 & 1.8 & 1.5 & \multicolumn{1}{c|}{0.1} & 16.7 & 1.9 & 21.9 & \multicolumn{1}{c|}{12.5} & 5.9 & 3.0 & \multicolumn{1}{c|}{5.2} & 1.2 & 1.4 & 2.6 & 1.2 & \multicolumn{1}{c|}{1.1} & 1 \\
\multicolumn{1}{l|}{BrightFlow} & \multicolumn{1}{c|}{3.2} & 2.4 & 0.7 & 4.9 & \multicolumn{1}{c|}{1.3} & 1.9 & 4.0 & 2.9 & \multicolumn{1}{c|}{1.1} & 10.2 & 5.4 & 20.7 & \multicolumn{1}{c|}{13.4} & 6.9 & 4.9 & \multicolumn{1}{c|}{6.3} & 2.1 & 2.4 & 2.6 & 2.0 & \multicolumn{1}{c|}{2.2} & 2 \\ \Xhline{3\arrayrulewidth}
\end{tabular}
}

\label{table:kitti_rcre}
\vskip -1.5\baselineskip plus -1fil
\end{table*}

\subsubsection{Supervised methods}

To ensure a fair and rigorous comparison across diverse network architectures, we standardize the downstream training pipeline for all evaluated supervised models while preserving their optimal base feature representations. We implement a unified experimental workflow inspired by the widely adopted protocol established in RAFT~\cite{teed2020raft}.

\noindent\textbf{OOD and ID Model Training.} 
We utilize the official pre-trained weights provided by the respective authors for the C+T stage (\textit{i.e.}, FlyingChairs~\cite{dosovitskiy2015flownet} and FlyingThings~\cite{mayer2016large}).
Building upon the C+T pre-trained weights, we strictly standardize the subsequent training stages for all models to establish our OOD and ID benchmarks:
1) \textit{OOD Setup:} We initialize each model with its C+T weights and train them on a unified mixed dataset comprising FlyingThings, Sintel~\cite{butler2012naturalistic}, and HD1K~\cite{kondermann2016hci} (C+T+S+H). Crucially, KITTI data is strictly excluded from this stage. This ensures that all models receive identical data exposure before testing and that the KITTI-FC test set remains entirely unseen, thereby rigorously assessing zero-shot robustness to outdoor driving domains.
2) \textit{ID Setup:} To evaluate in-domain adaptation, we utilize the fully trained OOD models as the base and fine-tune them exclusively on the clean versions of the 120 image pairs in the KITTI-FC training split, with standard training augmentations and without benchmark-corrupted training images.

\noindent\textbf{Training Hyperparameters.} 
We evaluated 12 supervised methods, including the recent SEA-RAFT~\cite{sea_raft} baseline; SAMFlow~\cite{zhou2024samflow} is assessed across three variants (SAMFlow-T, SAMFlow-B, and SAMFlow-H) using different SAM~\cite{kirillov2023segment} encoders. The supervised ID stage follows each method's KITTI fine-tuning recipe, with the configured learning rates, optimization budgets, effective batch sizes, and crop sizes provided in Supplementary Material, Sec.~S3 and Table~S3. We use AdamW with a weight decay of $10^{-5}$. For FlowDiffuser~\cite{luo2024flowdiffuser}, we adjusted the learning rate during OOD training to prevent training collapse.

\subsubsection{Unsupervised methods}

A primary advantage of unsupervised methods is their ability to directly adapt to the target domain using readily available unlabeled data. Therefore, evaluating them in an OOD setting does not reflect their optimal practical deployment strategy. Consequently, we focus our evaluation of unsupervised models exclusively on the ID benchmark of KITTI-FC. For the methods evaluated in our study, BrightFlow~\cite{marsal2023brightflow} and SMURF~\cite{stone2021smurf}, we directly utilize their official KITTI-trained checkpoints and default inference configurations to ensure a fair and rigorous assessment.

\subsection{Reporting Protocol and Ranking}
\label{sec:reporting_protocol}
For KITTI-FC, all $33$ variants are evaluated, with supervised methods reported under both OOD and ID. 
For GoPro-FC, only OOD models are considered (Sec.~\ref{sec:benchmarks}). 
As discussed in Sec.~\ref{sec:metrics}, different corruptions contribute unequally to the aggregate robustness due to diverse simulation mechanisms. 
Following the normalization strategy of~\cite{hendrycks2019benchmarking}, we employ a reference model to normalize corruption-wise results and then aggregate across corruptions to obtain a fair overall robustness score and ranking. 
We select RAFT-OOD as the reference because RAFT underpins many modern designs and OOD conditions are of primary interest in practice.

\begin{table*}[!t]
\centering
\caption{\textbf{$R^{d}$ results on KITTI-FC.} HL: \textit{High-light}, LL: \textit{Low-light}, OE: \textit{Over Exposure}, UE: \textit{Under Exposure}.}
\resizebox{\linewidth}{!}{
\begin{tabular}{lcccccccccccccccccccccc}
\Xhline{3\arrayrulewidth}
\multicolumn{1}{l|}{\multirow{2}{*}{\raisebox{-4ex}{\textbf{Model}}}} & \multicolumn{1}{c|}{\multirow{2}{*}{\raisebox{-4ex}{\textbf{Clean}}}} & \multicolumn{4}{c|}{\textbf{Digital}} & \multicolumn{4}{c|}{\textbf{Illumination}} & \multicolumn{4}{c|}{\textbf{Weather}} & \multicolumn{3}{c|}{\textbf{Noise}} & \multicolumn{5}{c|}{\textbf{Blur}} & \multicolumn{1}{c}{\multirow{2}{*}{\rotatebox{90}{RANK~~~~}}} \\ \cline{3-22} 
\multicolumn{1}{l|}{} & \multicolumn{1}{c|}{} & \rotatebox{90}{JPEG} & \rotatebox{90}{Pixelate} & \rotatebox{90}{Contrast} & \multicolumn{1}{c|}{\rotatebox{90}{Saturate}} & \rotatebox{90}{HL\tnote{*}} & \rotatebox{90}{LL\tnote{*}} & \rotatebox{90}{OE\tnote{*}} & \multicolumn{1}{c|}{\rotatebox{90}{UE\tnote{*}}} & \rotatebox{90}{Spatter} & \rotatebox{90}{Fog} & \rotatebox{90}{Frost} & \multicolumn{1}{c|}{\rotatebox{90}{Snow}} & \rotatebox{90}{Gaussian} & \rotatebox{90}{Shot} & \multicolumn{1}{c|}{\rotatebox{90}{Impulse}} & \rotatebox{90}{Gaussian} & \rotatebox{90}{Defocus} & \rotatebox{90}{Glass} & \rotatebox{90}{Camera} & \multicolumn{1}{c|}{\rotatebox{90}{PSF}}\\ \Xhline{3\arrayrulewidth}
\multicolumn{22}{c}{\textit{Knowledge-driven models}} \\ \Xhline{3\arrayrulewidth}
\multicolumn{1}{l|}{\textit{Farneb{\"a}ck}} & \multicolumn{1}{c|}{25.6} & 0.4 & 1.1 & 7.2 & \multicolumn{1}{c|}{-0.3} & 0.2 & 1.9 & 2.0 & \multicolumn{1}{c|}{3.0} & 2.2 & 5.6 & 4.1 & \multicolumn{1}{c|}{2.0} & -1.3 & -1.3 & \multicolumn{1}{c|}{-1.4} & 5.4 & 5.8 & 3.2 & 3.2 & \multicolumn{1}{c|}{5.0} & 2 \\
\multicolumn{1}{l|}{EpicFlow} & \multicolumn{1}{c|}{9.4} & 4.8 & 0.5 & 2.7 & \multicolumn{1}{c|}{2.2} & 1.5 & 4.1 & 6.3 & \multicolumn{1}{c|}{3.2} & 15.2 & 3.0 & 19.8 & \multicolumn{1}{c|}{10.6} & 17.8 & 23.9 & \multicolumn{1}{c|}{16.3} & 0.7 & 1.0 & 3.1 & 0.5 & \multicolumn{1}{c|}{0.5} & 1 \\
\multicolumn{1}{l|}{DIS} & \multicolumn{1}{c|}{20.6} & 0.5 & -0.1 & 0.3 & \multicolumn{1}{c|}{-0.2} & -0.5 & 1.4 & 3.8 & \multicolumn{1}{c|}{6.7} & 3.8 & 4.0 & 6.7 & \multicolumn{1}{c|}{2.6} & 0.8 & 0.5 & \multicolumn{1}{c|}{0.9} & -0.4 & -0.4 & -0.2 & -0.2 & \multicolumn{1}{c|}{-0.4} & 3 \\ \Xhline{3\arrayrulewidth}
\multicolumn{22}{c}{\textit{Out-Of-Domain (OOD) Supervised Models}} \\ \Xhline{3\arrayrulewidth}
\multicolumn{1}{l|}{RAFT} & \multicolumn{1}{c|}{4.3} & 6.0 & 0.2 & 0.5 & \multicolumn{1}{c|}{1.4} & 0.9 & 3.0 & 2.5 & \multicolumn{1}{c|}{0.1} & 17.2 & 2.1 & 23.5 & \multicolumn{1}{c|}{9.7} & 9.5 & 7.9 & \multicolumn{1}{c|}{9.9} & 1.9 & 2.0 & 4.7 & 0.9 & \multicolumn{1}{c|}{1.1} & 7 \\
\multicolumn{1}{l|}{GMA} & \multicolumn{1}{c|}{4.2} & 5.8 & 0.5 & 0.5 & \multicolumn{1}{c|}{1.5} & 0.8 & 3.2 & 2.0 & \multicolumn{1}{c|}{0.0} & 19.2 & 1.9 & 24.8 & \multicolumn{1}{c|}{10.0} & 10.8 & 9.4 & \multicolumn{1}{c|}{11.8} & 2.2 & 2.8 & 6.2 & 1.1 & \multicolumn{1}{c|}{1.3} & 9 \\
\multicolumn{1}{l|}{CSFlow} & \multicolumn{1}{c|}{4.1} & 5.1 & 0.3 & 0.7 & \multicolumn{1}{c|}{1.0} & 0.8 & 3.4 & 2.4 & \multicolumn{1}{c|}{0.0} & 17.4 & 2.1 & 23.4 & \multicolumn{1}{c|}{9.9} & 6.7 & 5.6 & \multicolumn{1}{c|}{6.9} & 1.5 & 1.8 & 4.5 & 1.0 & \multicolumn{1}{c|}{1.0} & 2 \\
\multicolumn{1}{l|}{SKFlow} & \multicolumn{1}{c|}{4.0} & 4.9 & 0.2 & 0.6 & \multicolumn{1}{c|}{1.3} & 0.6 & 2.7 & 1.7 & \multicolumn{1}{c|}{0.0} & 16.8 & 1.6 & 22.6 & \multicolumn{1}{c|}{9.7} & 10.7 & 9.6 & \multicolumn{1}{c|}{11.6} & 2.1 & 2.5 & 5.2 & 1.1 & \multicolumn{1}{c|}{1.3} & 5 \\
\multicolumn{1}{l|}{GMFlowNet} & \multicolumn{1}{c|}{4.1} & 5.3 & 0.4 & 0.8 & \multicolumn{1}{c|}{1.2} & 0.9 & 4.0 & 1.9 & \multicolumn{1}{c|}{0.2} & 17.4 & 2.3 & 23.0 & \multicolumn{1}{c|}{7.5} & 8.4 & 7.2 & \multicolumn{1}{c|}{9.0} & 1.5 & 1.8 & 4.1 & 1.0 & \multicolumn{1}{c|}{0.9} & 8 \\
\multicolumn{1}{l|}{CRAFT} & \multicolumn{1}{c|}{4.7} & 5.8 & 0.5 & 0.9 & \multicolumn{1}{c|}{1.9} & 0.9 & 3.3 & 2.4 & \multicolumn{1}{c|}{0.1} & 18.5 & 2.2 & 25.5 & \multicolumn{1}{c|}{13.9} & 12.5 & 11.0 & \multicolumn{1}{c|}{13.2} & 2.3 & 2.7 & 7.6 & 1.5 & \multicolumn{1}{c|}{1.3} & 12 \\
\multicolumn{1}{l|}{FlowFormer} & \multicolumn{1}{c|}{4.8} & 5.1 & 0.5 & 2.4 & \multicolumn{1}{c|}{1.6} & 0.9 & 3.7 & 1.9 & \multicolumn{1}{c|}{-0.1} & 18.5 & 2.3 & 24.2 & \multicolumn{1}{c|}{10.3} & 8.5 & 7.2 & \multicolumn{1}{c|}{9.4} & 2.1 & 2.0 & 5.7 & 1.1 & \multicolumn{1}{c|}{1.2} & 10 \\
\multicolumn{1}{l|}{FlowFormer++} & \multicolumn{1}{c|}{4.7} & 6.0 & 1.2 & 1.8 & \multicolumn{1}{c|}{2.0} & 1.3 & 4.1 & 1.9 & \multicolumn{1}{c|}{0.1} & 18.7 & 2.1 & 24.7 & \multicolumn{1}{c|}{9.8} & 9.3 & 8.4 & \multicolumn{1}{c|}{10.3} & 2.0 & 2.1 & 6.8 & 1.4 & \multicolumn{1}{c|}{1.4} & 13 \\
\multicolumn{1}{l|}{SAMFlow-T} & \multicolumn{1}{c|}{4.2} & 5.0 & 0.5 & 1.5 & \multicolumn{1}{c|}{1.3} & 0.9 & 2.4 & 1.7 & \multicolumn{1}{c|}{-0.1} & 17.4 & 1.5 & 23.4 & \multicolumn{1}{c|}{8.6} & 9.0 & 7.7 & \multicolumn{1}{c|}{8.3} & 1.9 & 2.0 & 4.8 & 0.6 & \multicolumn{1}{c|}{1.0} & 6 \\
\multicolumn{1}{l|}{SAMFlow-B} & \multicolumn{1}{c|}{4.1} & 4.0 & 0.4 & 1.0 & \multicolumn{1}{c|}{1.2} & 0.9 & 2.3 & 1.7 & \multicolumn{1}{c|}{0.0} & 17.4 & 1.4 & 23.4 & \multicolumn{1}{c|}{7.1} & 6.6 & 5.8 & \multicolumn{1}{c|}{6.2} & 1.6 & 1.9 & 4.9 & 0.6 & \multicolumn{1}{c|}{0.5} & 1 \\
\multicolumn{1}{l|}{SAMFlow-H} & \multicolumn{1}{c|}{3.8} & 3.3 & 0.4 & 1.2 & \multicolumn{1}{c|}{1.6} & 1.0 & 2.5 & 1.7 & \multicolumn{1}{c|}{0.0} & 17.4 & 1.5 & 23.5 & \multicolumn{1}{c|}{7.0} & 8.0 & 6.8 & \multicolumn{1}{c|}{6.8} & 1.6 & 1.8 & 4.7 & 0.4 & \multicolumn{1}{c|}{1.0} & 4 \\
\multicolumn{1}{l|}{FlowDiffuser} & \multicolumn{1}{c|}{6.2} & 8.0 & 0.5 & 1.7 & \multicolumn{1}{c|}{2.3} & 2.1 & 3.3 & 3.1 & \multicolumn{1}{c|}{0.0} & 16.3 & 1.9 & 23.0 & \multicolumn{1}{c|}{16.6} & 13.2 & 11.4 & \multicolumn{1}{c|}{13.7} & 3.2 & 3.3 & 5.6 & 2.2 & \multicolumn{1}{c|}{1.7} & 14 \\
\multicolumn{1}{l|}{DPFlow} & \multicolumn{1}{c|}{3.8} & 4.7 & 0.2 & 1.0 & \multicolumn{1}{c|}{0.5} & 0.4 & 2.2 & 0.7 & \multicolumn{1}{c|}{-0.1} & 16.7 & 2.7 & 22.9 & \multicolumn{1}{c|}{9.8} & 5.0 & 3.3 & \multicolumn{1}{c|}{5.3} & 4.8 & 3.2 & 5.6 & 1.0 & \multicolumn{1}{c|}{1.9} & 3 \\
\multicolumn{1}{l|}{SEA-RAFT} & \multicolumn{1}{c|}{3.8} & 6.7 & 0.6 & 0.9 & \multicolumn{1}{c|}{1.0} & 0.9 & 2.5 & 2.1 & \multicolumn{1}{c|}{0.2} & 17.2 & 2.8 & 22.9 & \multicolumn{1}{c|}{12.1} & 7.2 & 5.7 & \multicolumn{1}{c|}{7.8} & 2.4 & 2.8 & 6.9 & 1.1 & \multicolumn{1}{c|}{1.6} & 11 \\ \Xhline{3\arrayrulewidth}
\multicolumn{22}{c}{\textit{In-Domain (ID) Supervised Models}} \\ \Xhline{3\arrayrulewidth}
\multicolumn{1}{l|}{RAFT} & \multicolumn{1}{c|}{1.8} & 3.8 & 0.3 & 2.5 & \multicolumn{1}{c|}{0.7} & 1.5 & 5.0 & 5.8 & \multicolumn{1}{c|}{0.1} & 10.2 & 1.5 & 18.6 & \multicolumn{1}{c|}{34.0} & 14.2 & 9.1 & \multicolumn{1}{c|}{13.0} & 2.7 & 2.5 & 2.7 & 3.6 & \multicolumn{1}{c|}{2.4} & 13 \\
\multicolumn{1}{l|}{GMA} & \multicolumn{1}{c|}{2.4} & 3.3 & 0.8 & 1.1 & \multicolumn{1}{c|}{0.6} & 3.9 & 2.2 & 5.1 & \multicolumn{1}{c|}{-0.2} & 12.6 & 1.9 & 20.6 & \multicolumn{1}{c|}{49.3} & 20.1 & 12.8 & \multicolumn{1}{c|}{18.9} & 1.8 & 1.9 & 2.5 & 3.7 & \multicolumn{1}{c|}{1.4} & 11 \\
\multicolumn{1}{l|}{CSFlow} & \multicolumn{1}{c|}{2.0} & 1.8 & 0.3 & 1.2 & \multicolumn{1}{c|}{0.7} & 0.6 & 3.1 & 2.1 & \multicolumn{1}{c|}{0.1} & 9.2 & 1.6 & 17.4 & \multicolumn{1}{c|}{15.4} & 9.3 & 6.5 & \multicolumn{1}{c|}{11.0} & 1.3 & 1.3 & 1.5 & 1.0 & \multicolumn{1}{c|}{0.8} & 4 \\
\multicolumn{1}{l|}{SKFlow} & \multicolumn{1}{c|}{2.8} & 1.2 & -0.2 & 0.5 & \multicolumn{1}{c|}{0.7} & 2.4 & 1.1 & 4.4 & \multicolumn{1}{c|}{0.0} & 10.5 & 0.8 & 17.6 & \multicolumn{1}{c|}{36.9} & 20.9 & 14.3 & \multicolumn{1}{c|}{22.7} & 0.4 & 0.5 & 1.3 & 1.2 & \multicolumn{1}{c|}{0.0} & 6 \\
\multicolumn{1}{l|}{GMFlowNet} & \multicolumn{1}{c|}{1.8} & 1.6 & 0.2 & 1.0 & \multicolumn{1}{c|}{0.3} & 0.9 & 2.3 & 6.9 & \multicolumn{1}{c|}{0.1} & 10.4 & 1.3 & 18.4 & \multicolumn{1}{c|}{20.7} & 13.2 & 10.9 & \multicolumn{1}{c|}{14.5} & 1.5 & 1.5 & 1.8 & 1.3 & \multicolumn{1}{c|}{1.1} & 7 \\
\multicolumn{1}{l|}{CRAFT} & \multicolumn{1}{c|}{3.7} & 2.8 & 0.9 & 0.7 & \multicolumn{1}{c|}{1.7} & 1.2 & 1.1 & 2.1 & \multicolumn{1}{c|}{0.6} & 10.2 & -0.1 & 19.5 & \multicolumn{1}{c|}{28.3} & 11.1 & 8.2 & \multicolumn{1}{c|}{11.1} & 1.8 & 2.0 & 4.0 & 1.3 & \multicolumn{1}{c|}{1.0} & 12 \\
\multicolumn{1}{l|}{FlowFormer} & \multicolumn{1}{c|}{1.7} & 2.0 & 0.6 & 2.9 & \multicolumn{1}{c|}{0.5} & 1.0 & 2.7 & 2.1 & \multicolumn{1}{c|}{0.2} & 14.6 & 2.2 & 20.9 & \multicolumn{1}{c|}{20.2} & 7.1 & 5.1 & \multicolumn{1}{c|}{7.8} & 2.7 & 2.7 & 2.3 & 1.9 & \multicolumn{1}{c|}{1.8} & 10 \\
\multicolumn{1}{l|}{FlowFormer++} & \multicolumn{1}{c|}{1.6} & 2.2 & 0.8 & 1.8 & \multicolumn{1}{c|}{0.5} & 0.6 & 2.6 & 1.3 & \multicolumn{1}{c|}{0.0} & 15.4 & 1.5 & 22.1 & \multicolumn{1}{c|}{11.0} & 8.0 & 5.9 & \multicolumn{1}{c|}{7.5} & 2.0 & 2.1 & 2.6 & 1.4 & \multicolumn{1}{c|}{1.5} & 8 \\
\multicolumn{1}{l|}{SAMFlow-T} & \multicolumn{1}{c|}{1.5} & 1.7 & 0.5 & 2.2 & \multicolumn{1}{c|}{0.3} & 0.4 & 1.8 & 0.6 & \multicolumn{1}{c|}{0.1} & 15.3 & 1.3 & 20.6 & \multicolumn{1}{c|}{9.7} & 3.6 & 2.9 & \multicolumn{1}{c|}{3.8} & 1.4 & 1.4 & 1.9 & 1.2 & \multicolumn{1}{c|}{1.1} & 5 \\
\multicolumn{1}{l|}{SAMFlow-B} & \multicolumn{1}{c|}{1.5} & 1.8 & 0.5 & 1.6 & \multicolumn{1}{c|}{0.3} & 0.8 & 1.7 & 0.6 & \multicolumn{1}{c|}{0.0} & 16.0 & 1.1 & 21.0 & \multicolumn{1}{c|}{8.2} & 4.3 & 3.1 & \multicolumn{1}{c|}{4.0} & 1.1 & 1.1 & 2.4 & 1.2 & \multicolumn{1}{c|}{0.8} & 3 \\
\multicolumn{1}{l|}{SAMFlow-H} & \multicolumn{1}{c|}{1.5} & 1.4 & 0.5 & 1.1 & \multicolumn{1}{c|}{0.3} & 0.4 & 1.5 & 0.6 & \multicolumn{1}{c|}{0.1} & 15.4 & 0.9 & 20.3 & \multicolumn{1}{c|}{9.1} & 3.2 & 1.9 & \multicolumn{1}{c|}{3.3} & 1.0 & 1.1 & 2.0 & 0.9 & \multicolumn{1}{c|}{0.6} & 2 \\
\multicolumn{1}{l|}{FlowDiffuser} & \multicolumn{1}{c|}{4.7} & 4.1 & 0.2 & 3.3 & \multicolumn{1}{c|}{2.3} & 2.5 & 6.1 & 5.0 & \multicolumn{1}{c|}{0.9} & 12.1 & 2.0 & 19.6 & \multicolumn{1}{c|}{27.6} & 23.5 & 20.2 & \multicolumn{1}{c|}{23.4} & 1.3 & 1.4 & 3.9 & 0.8 & \multicolumn{1}{c|}{0.8} & 14 \\
\multicolumn{1}{l|}{DPFlow} & \multicolumn{1}{c|}{1.5} & 1.2 & 0.2 & 0.4 & \multicolumn{1}{c|}{0.1} & 0.2 & 2.5 & 0.4 & \multicolumn{1}{c|}{0.0} & 11.0 & 1.2 & 16.8 & \multicolumn{1}{c|}{10.1} & 2.5 & 0.9 & \multicolumn{1}{c|}{2.0} & 1.0 & 1.0 & 1.3 & 1.0 & \multicolumn{1}{c|}{0.8} & 1 \\
\multicolumn{1}{l|}{SEA-RAFT} & \multicolumn{1}{c|}{1.7} & 3.1 & 0.2 & 0.9 & \multicolumn{1}{c|}{0.2} & 0.7 & 2.1 & 2.2 & \multicolumn{1}{c|}{0.2} & 9.1 & 1.1 & 16.3 & \multicolumn{1}{c|}{42.5} & 2.4 & 1.7 & \multicolumn{1}{c|}{2.5} & 1.4 & 1.5 & 1.7 & 4.6 & \multicolumn{1}{c|}{1.2} & 9 \\ \Xhline{3\arrayrulewidth}
\multicolumn{22}{c}{\textit{In-Domain (ID) Unsupervised Models}} \\ \Xhline{3\arrayrulewidth}
\multicolumn{1}{l|}{SMURF} & \multicolumn{1}{c|}{2.1} & 1.1 & 0.1 & 0.0 & \multicolumn{1}{c|}{0.1} & 0.2 & 1.0 & 1.1 & \multicolumn{1}{c|}{0.0} & 15.7 & 1.3 & 20.6 & \multicolumn{1}{c|}{11.4} & 4.9 & 2.2 & \multicolumn{1}{c|}{4.2} & 0.5 & 0.6 & 1.8 & 0.4 & \multicolumn{1}{c|}{0.3} & 1 \\
\multicolumn{1}{l|}{BrightFlow} & \multicolumn{1}{c|}{3.2} & 1.7 & 0.3 & 3.7 & \multicolumn{1}{c|}{0.6} & 1.0 & 2.8 & 1.7 & \multicolumn{1}{c|}{0.5} & 9.1 & 4.3 & 19.1 & \multicolumn{1}{c|}{11.3} & 5.4 & 3.6 & \multicolumn{1}{c|}{4.7} & 1.3 & 1.4 & 1.6 & 1.0 & \multicolumn{1}{c|}{1.2} & 2 \\
\Xhline{3\arrayrulewidth}
\end{tabular}
}
\label{table:kitti_cre}
\end{table*}

\subsection{Overall Results}
\label{sec:overall_results}
We report $R^{s}$ and $R^{d}$ on KITTI-FC (OOD/ID), and the GT-free $R^{s}$ on GoPro-FC (10/30 FPS). 
Comprehensive tables are provided in Tab.~\ref{table:kitti_rcre}, Tab.~\ref{table:kitti_cre}, and Tab.~\ref{table:gopro-fc}, with additional analyses across severity levels and corruption types in Tab.~\ref{tab:HL_LL_OE_UE}, Tab.~\ref{tab:JPEG_Sat_Spa_Fog}, and Tab.~\ref{tab:UE_Pixelate_across_severity}. 
Supplementary Material, Secs.~S1--S2, provides additional analyses of cross-frame corruption effects on KITTI-FC and motion-scale normalization on GoPro-FC.
The subsequent subsections turn from reporting to interpretation: first examining metric behavior, then analyzing corruption influence, and finally discussing model capability.

\subsection{Metric Analysis and Relationships}

We characterize robustness along two complementary axes. $R^{s}$ measures the output shift caused by corruption, whereas $R^{d}$ measures the signed change in ground-truth accuracy. Their clean-error-normalized forms, $R^{s}_{\mathrm{rel}}$ and $R^{d}_{\mathrm{rel}}$, express these effects relative to each model's clean EPE. Figure~\ref{fig:metric_chart} visualizes pairwise relationships after reference normalization.

Clean EPE is positively associated with the two unnormalized corruption responses for many evaluated learning-based models, but the relationship is neither exact nor universal. Moreover, the relative metrics can change substantially after clean-domain fine-tuning even when clean EPE improves. These observations support reporting stability and accuracy degradation separately instead of treating clean accuracy as a proxy for robustness. Because the evaluated checkpoints differ in architecture, training data, and optimization, these plots support descriptive associations rather than causal conclusions about any single design choice.

\begin{table*}[!t]
\centering
\setlength{\tabcolsep}{2pt}
\begin{threeparttable}

\caption{\textbf{$R^{s}$ results on GoPro-FC at 30/10 FPS.} FF: FlowFormer, FF++: FlowFormer++.}
\begin{tabular}{ll|ccccccccc}
\Xhline{3\arrayrulewidth}
\multicolumn{2}{c|}{Corruption} & RAFT & GMA & CSFlow & SKFlow & GMFlowNet & CRAFT & FF\tnote{**} & FF++\tnote{**} & SAMFlow-H \\ \Xhline{3\arrayrulewidth}
\multirow{4}{*}{\rotatebox{90}{Digital}} & JPEG & 1.3 / 3.1 & 1.2 / 2.3 & 1.4 / 2.8 & 0.8 / 1.7 & 1.4 / 3.0 & 0.8 / 1.7 & 1.6 / 2.7 & 1.3 / 2.9 & 1.3 / 2.7 \\
 & Pixelate & 0.2 / 0.6 & 0.2 / 0.6 & 0.2 / 0.5 & 0.2 / 0.5 & 0.2 / 0.6 & 0.2 / 0.6 & 0.3 / 0.7 & 0.4 / 0.9 & 0.4 / 0.7 \\
 & Contrast & 0.3 / 0.9 & 0.3 / 0.8 & 0.3 / 0.8 & 0.2 / 0.7 & 0.2 / 0.7 & 0.3 / 1.0 & 0.3 / 0.9 & 0.4 / 0.9 & 0.4 / 0.8 \\
 & Saturate & 0.4 / 1.4 & 0.3 / 0.8 & 0.5 / 1.3 & 0.3 / 0.9 & 0.4 / 1.2 & 0.3 / 0.9 & 0.4 / 1.0 & 0.4 / 1.0 & 0.4 / 1.4 \\ \hline
\multirow{4}{*}{\rotatebox{90}{Illu.}} & HL\tnote{*} & 0.2 / 0.8 & 0.2 / 0.6 & 0.2 / 0.7 & 0.2 / 0.6 & 0.2 / 0.7 & 0.2 / 0.6 & 0.2 / 0.6 & 0.2 / 0.6 & 0.2 / 0.6 \\
 & LL\tnote{*} & 0.4 / 1.3 & 0.3 / 1.0 & 0.4 / 1.5 & 0.3 / 1.0 & 0.5 / 1.7 & 0.4 / 1.1 & 0.4 / 1.1 & 0.4 / 1.2 & 0.4 / 1.0 \\
 & OE\tnote{*} & 0.7 / 1.8 & 0.5 / 1.1 & 0.7 / 1.9 & 0.4 / 1.0 & 0.6 / 2.0 & 0.4 / 1.2 & 0.6 / 1.6 & 0.4 / 1.3 & 0.4 / 2.9 \\
 & UE\tnote{*} & 0.1 / 0.4 & 0.1 / 0.2 & 0.1 / 0.3 & 0.1 / 0.3 & 0.1 / 0.3 & 0.1 / 0.2 & 0.1 / 0.2 & 0.1 / 0.3 & 0.1 / 0.2 \\ \hline
\multirow{4}{*}{\rotatebox{90}{Weather}} & Spatter & 7.6 / 21.1 & 7.7 / 19.9 & 7.6 / 21.0 & 7.4 / 18.6 & 7.6 / 20.0 & 7.4 / 17.8 & 7.8 / 18.8 & 8.3 / 21.1 & 8.3 / 19.5 \\
 & Fog & 0.7 / 2.9 & 0.5 / 1.6 & 0.8 / 2.7 & 0.3 / 1.0 & 0.8 / 2.5 & 0.5 / 1.3 & 0.7 / 1.6 & 0.6 / 1.3 & 0.6 / 1.3 \\
 & Frost & 11.0 / 32.0 & 11.1 / 32.0 & 10.8 / 30.8 & 9.9 / 26.4 & 10.6 / 30.7 & 10.1 / 29.4 & 11.0 / 30.9 & 11.3 / 31.8 & 11.3 / 30.7 \\
 & Snow & 3.6 / 5.6 & 1.2 / 2.9 & 2.3 / 4.3 & 1.1 / 2.7 & 2.0 / 5.5 & 1.3 / 3.2 & 2.2 / 3.9 & 1.6 / 3.7 & 1.6 / 5.8 \\ \hline
\multirow{3}{*}{\rotatebox{90}{Noise}} & Gaussian & 1.4 / 3.6 & 1.4 / 3.9 & 1.6 / 3.8 & 1.4 / 3.5 & 1.3 / 4.1 & 1.3 / 3.3 & 0.9 / 2.4 & 1.1 / 3.3 & 1.1 / 3.2 \\
 & Shot & 1.3 / 3.4 & 1.3 / 3.4 & 1.4 / 3.3 & 1.3 / 3.5 & 1.3 / 3.7 & 1.3 / 3.2 & 0.9 / 2.5 & 1.1 / 3.1 & 1.1 / 2.9 \\
 & Impulse & 1.4 / 3.5 & 1.3 / 3.7 & 1.5 / 3.4 & 1.4 / 3.5 & 1.3 / 4.1 & 1.3 / 3.2 & 1.0 / 2.6 & 1.1 / 3.2 & 1.1 / 2.4 \\ \hline
\multirow{6}{*}{\rotatebox{90}{Blur}} & Gaussian & 0.4 / 1.8 & 0.4 / 1.0 & 0.4 / 1.4 & 0.5 / 1.4 & 0.4 / 1.0 & 0.4 / 1.3 & 0.4 / 1.0 & 0.4 / 1.0 & 0.4 / 0.9 \\
 & Defocus & 0.4 / 1.9 & 0.4 / 1.1 & 0.4 / 1.5 & 0.4 / 1.4 & 0.4 / 1.1 & 0.4 / 1.2 & 0.4 / 1.0 & 0.4 / 1.1 & 0.4 / 1.0 \\
 & Glass & 0.7 / 1.5 & 0.7 / 1.2 & 0.7 / 1.5 & 0.7 / 1.4 & 0.7 / 1.2 & 0.7 / 1.3 & 0.7 / 1.1 & 0.7 / 1.2 & 0.7 / 1.1 \\
 & Camera & 0.4 / 1.8 & 0.4 / 1.0 & 0.4 / 1.5 & 0.4 / 1.2 & 0.4 / 1.1 & 0.4 / 1.0 & 0.4 / 1.0 & 0.4 / 1.1 & 0.4 / 1.0 \\
 & Object & 1.2 / 3.1 & 1.1 / 2.0 & 1.2 / 2.7 & 1.0 / 2.0 & 1.1 / 2.4 & 1.1 / 2.0 & 1.0 / 1.8 & 1.0 / 2.4 & 1.0 / 2.1 \\
 & PSF & 0.4 / 1.4 & 0.3 / 1.0 & 0.3 / 1.4 & 0.4 / 1.2 & 0.3 / 1.0 & 0.4 / 1.1 & 0.3 / 1.0 & 0.3 / 0.9 & 0.3 / 0.9 \\ \hline
\multirow{3}{*}{\rotatebox{0}{Video}} & ABR & 1.7 / 2.9 & 1.5 / 2.6 & 1.7 / 3.2 & 1.4 / 2.6 & 1.6 / 3.1 & 1.5 / 2.5 & 1.8 / 3.2 & 1.7 / 3.3 & 1.7 / 3.4 \\
 & CRF & 2.6 / 5.2 & 2.4 / 4.1 & 2.7 / 5.3 & 2.3 / 4.1 & 2.6 / 5.3 & 2.3 / 3.9 & 2.8 / 5.2 & 2.7 / 5.4 & 2.7 / 5.8 \\
 & BitError & 0.5 / 1.4 & 0.3 / 1.0 & 0.4 / 1.2 & 0.3 / 0.9 & 0.6 / 1.9 & 0.4 / 0.9 & 0.4 / 1.1 & 0.4 / 1.0 & 0.4 / 1.5 \\ \hline
\multicolumn{2}{c|}{AVG RANK} & 9 / 8 & 3 / 2 & 8 / 7 & 1 / 1 & 7 / 6 & 2 / 3 & 4 / 4 & 5 / 5 & 6 / 9 \\\Xhline{3\arrayrulewidth}
\end{tabular}

\label{table:gopro-fc}

\end{threeparttable}
\vskip -1.5\baselineskip plus -1fil
\end{table*}

\begin{table}[t]
\centering
\caption{PSNR of four illumination corruptions. EPE, $R^{s}$, and $R^{d}$ for RAFT-OOD are also reported.}
\begin{tabular}{l|c|ccc}
\Xhline{3\arrayrulewidth}
\multirow{2}{*}{Corruption} & \multirow{2}{*}{PSNR} & \multicolumn{3}{c}{RAFT-OOD} \\ \cline{3-5} 
 &  & EPE & $R^{s}$ & $R^{d}$ \\ \hline
High-Light & 13.97 & 5.2 & 2.01 & 0.90 \\
Low-Light & 14.53 & 7.25 & 4.16 & 2.95 \\ \hline
Over Exposure & 13.32 & 6.74 & 3.30 & 2.45 \\
Under Exposure & 12.54 & 4.34 & 0.43 & 0.05 \\ \Xhline{3\arrayrulewidth}
\end{tabular}
\label{tab:HL_LL_OE_UE}

\vskip -1.5\baselineskip plus -1fil
\end{table}

\subsection{Corruption Influence}
\label{sec:corruption_influence}

To understand how different types of corruptions affect optical flow estimation, we conduct detailed analyses on the KITTI-FC and GoPro-FC benchmark, covering both quantitative performance and qualitative visual evidence. 
We summarize three key findings that reveal the intrinsic sensitivity of optical flow models to various corruption categories.

\noindent \textbf{Finding 1: } \textit{Low-Light (LL) and Over Exposure (OE) conditions are more detrimental than High-Light (HL) and Under Exposure (UE), respectively.}
Illumination corruptions share similar implementation patterns within each subgroup (\textit{e.g.}, HL vs. LL, OE vs. UE), allowing direct comparison. 
As summarized in Tab.~\ref{table:kitti_rcre}, Tab.~\ref{table:kitti_cre} and Tab.~\ref{table:gopro-fc}, all tested models exhibit a notable robustness drop in the \textit{Low-Light} setting compared with \textit{High-Light}. 
Similarly, \textit{Over Exposure} leads to more severe degradation than \textit{Under Exposure}. 
Interestingly, the PSNR values (Tab.~\ref{tab:HL_LL_OE_UE}) indicate that LL and OE images are visually closer to the clean ones, yet models perform worse on them. 
This apparent contradiction can be explained by the evaluation characteristics and pixel-wise illumination effects.

For the implementation of \textit{High-Light} and \textit{Low-Light}, we apply a global shift to the image luminance. 
Within the KITTI-FC and GoPro-FC datasets, the mean luminance of pixels registers approximately between 102 and 107, which falls short of the midpoint value of 128 for 8-bit images. 
Consequently, the proportion of low-luminance pixels affected by \textit{Low-Light} conditions exceeds that of high-luminance pixels impacted by \textit{High-Light} scenarios, thereby signifying the greater severity of \textit{Low-Light} corruptions.
Regarding exposure, the brightness modification in OE and UE is nonlinear and dependent on pixel intensity. 
In photography, judicious under-exposure frequently facilitates the capture of a broader dynamic range~\cite{hasinoff2016burst}.
\textit{Over Exposure} causes irreversible saturation in bright areas, thereby losing local contrast and structural details critical for correspondence, while \textit{Under Exposure} preserves more contrast information, explaining the performance asymmetry.

\begin{figure}
    \centering
    \includegraphics[width=0.85\linewidth]{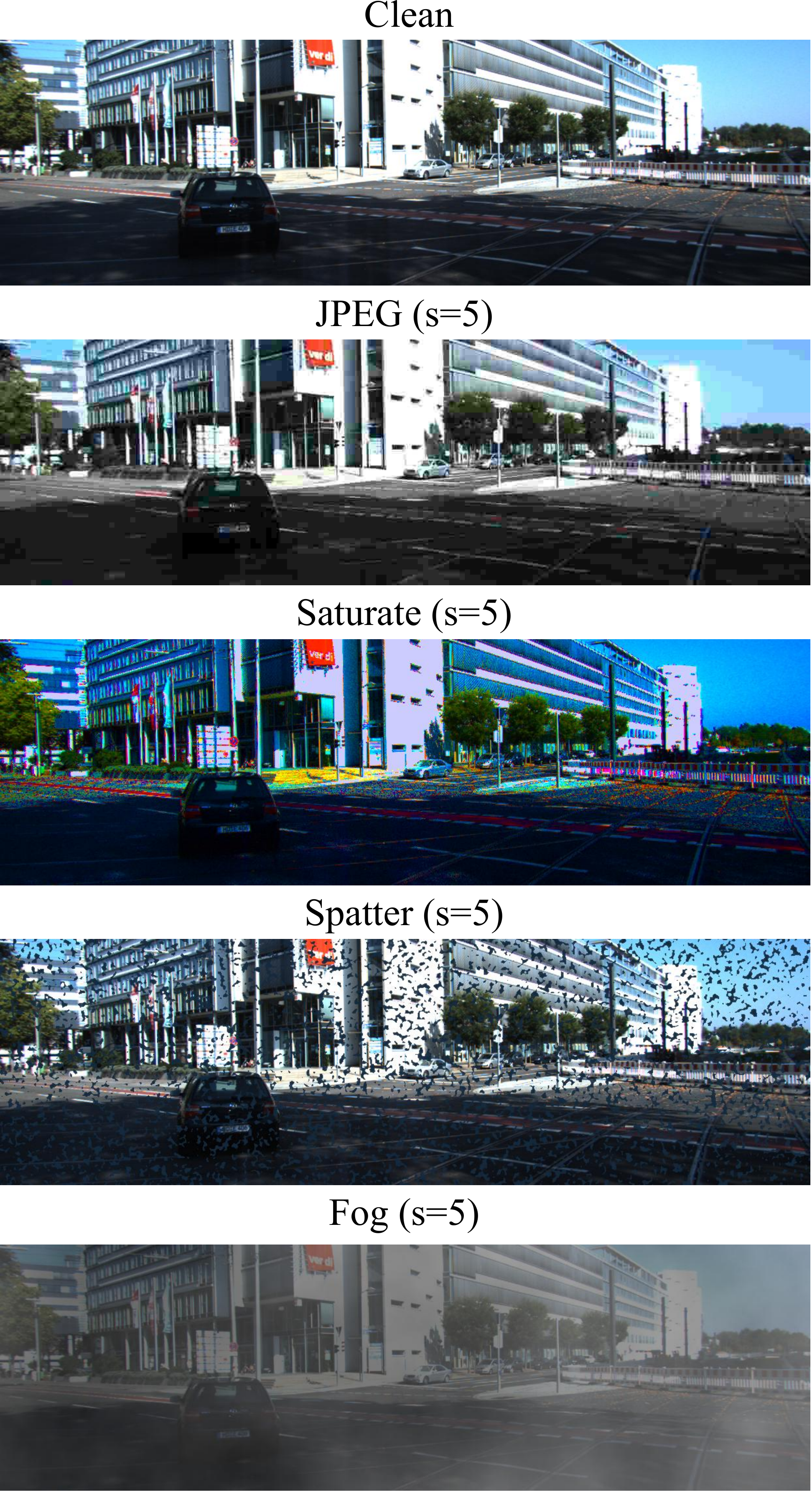}
    \caption{\textbf{JPEG, Saturate, Spatter, and Fog samples from KITTI-FC.} 
    JPEG removes local high-frequency cues, while Saturate modifies color distribution. 
    Spatter introduces occlusions, and Fog lowers contrast while retaining local structure.}
    \label{fig:JPEG_Sat_Spa_Fog}
    \vskip -1.5\baselineskip plus -1fil
\end{figure}

\noindent \textbf{Finding 2: } \textit{Under the evaluated corruption schedules, local structural damage can be more harmful than global appearance changes.}
JPEG, Noise, and occlusion-like effects remove fine local cues required for correspondence, whereas Contrast, Saturate, and Fog primarily change global appearance while retaining local gradients. Accordingly, Fig.~\ref{fig:JPEG_Sat_Spa_Fog} and Tab.~\ref{tab:JPEG_Sat_Spa_Fog} show that the visually subtler JPEG and Spatter corruptions cause larger errors than Saturate and Fog. SEA-RAFT follows the same KITTI-FC trend: its OOD $R^{s}$ is $8.4/18.4$ for JPEG/Spatter versus $1.9/3.7$ for Saturate/Fog (Tab.~\ref{table:kitti_rcre}).

The above comparisons use corruption-specific severity schedules rather than a common image-quality scale. To assess whether the trend persists at comparable image quality, we further evaluate JPEG and Spatter against Saturate using RAFT-OOD, restricting PSNR and SSIM differences to at most 1 dB and 0.03. Both corruptions still produce larger output shifts and accuracy degradation than Saturate, supporting the importance of local correspondence cues within the tested quality range. Details and quantitative results are provided in Supplementary Material, Sec.~S4 and Table~S4.

\noindent \textbf{Finding 3: } \textit{Certain corruptions can paradoxically improve estimation performance.}
As shown in Tab.~\ref{tab:UE_Pixelate_across_severity}, the accuracy degradation $R^{d}$ becomes negative under specific conditions such as \textit{Under Exposure} or \textit{Pixelate}. 
This implies that the model performs better on the corrupted input than on the clean data, challenging the conventional assumption that corruption universally degrades performance. 
Furthermore, certain perturbations applied to images, such as denoising, can indeed be demonstrated to aid in optical flow estimation~\cite{yu2020joint}.
We hypothesize that this improvement occurs when corruptions inadvertently suppress ``shortcut'' features to which models have overfitted during training, such as sensor noise, excessive texture detail, or irrelevant local patterns. 
By filtering these misleading cues, the corrupted data encourage models to rely more on robust, global motion structures, thereby improving accuracy. 
This observation opens an interesting direction for future research in test-time data augmentation or corruption-guided regularization to enhance optical flow robustness.

Overall, these analyses reveal that optical flow models are more vulnerable to information loss at the local structural level than to global perceptual distortion under the evaluated schedules, and that certain mild corruptions may even expose beneficial inductive biases. 
Such insights not only deepen our understanding of model robustness but also point toward new strategies for adaptive testing and model design in real-world deployment scenarios.

\begin{table}[t]
\centering
\caption{PSNR of four representative corruptions. JPEG and Spatter have higher PSNR but cause larger optical-flow errors because they remove local information or introduce occlusion. EPE, $R^{s}$, and $R^{d}$ for RAFT-OOD are also reported.}
\begin{tabular}{l|c|clc}
\Xhline{3\arrayrulewidth}
\multirow{2}{*}{Corruption} & \multicolumn{1}{c|}{\multirow{2}{*}{PSNR}} & \multicolumn{3}{c}{RAFT-OOD} \\ \cline{3-5} 
 & \multicolumn{1}{c|}{} & \multicolumn{1}{c}{EPE} & \multicolumn{1}{c}{$R^{s}$} & \multicolumn{1}{c}{$R^{d}$} \\ \hline
JPEG & 25.76 & 10.28 & 7.57 & 5.99 \\
Saturate & 19.29 & 5.73 & 2.41 & 1.44 \\ \hline
Spatter & 24.78 & 21.49 & 18.59 & 17.2 \\
Fog & 11.69 & 6.34 & 3.14 & 2.05 \\ \Xhline{3\arrayrulewidth}
\end{tabular}
\label{tab:JPEG_Sat_Spa_Fog}

\end{table}

\begin{table*}[t]
\centering

\caption{\textbf{$R^{d}$ across severity levels for Under Exposure and Pixelate.}}
\begin{tabular}{c|c|ccccc|ccccc}
\Xhline{3\arrayrulewidth}
\multirow{2}{*}{Setting} & \multirow{2}{*}{Method} & \multicolumn{5}{c|}{UE} & \multicolumn{5}{c}{Pixelate} \\ \cline{3-12} 
 &  & 1 & 2 & 3 & 4 & 5 & 1 & 2 & 3 & 4 & 5 \\ \hline
\multirow{2}{*}{OOD} & RAFT & -0.002 & 0.010 & 0.095 & 0.140 & 0.011 & 0.104 & 0.055 & 0.262 & 0.411 & 0.363 \\
 & GMA & -0.044 & -0.001 & 0.010 & 0.087 & -0.033 & -0.061 & 0.009 & 0.175 & 1.263 & 1.147 \\ \hline
\multirow{2}{*}{ID} & RAFT & 0.059 & 0.520 & 1.092 & 2.569 & 4.173 & -0.028 & -0.026 & 0.168 & 0.447 & 0.628 \\
 & GMA & -0.138 & -0.207 & -0.054 & -0.295 & -0.041 & -0.190 & 0.187 & 0.723 & 1.482 & 1.846 \\ \Xhline{3\arrayrulewidth}
\end{tabular}
\label{tab:UE_Pixelate_across_severity}
\end{table*}
\subsection{Model-Property Analysis}
\label{sec:model_capability}

\begin{figure}[t]
    \centering
    \includegraphics[width=\linewidth]{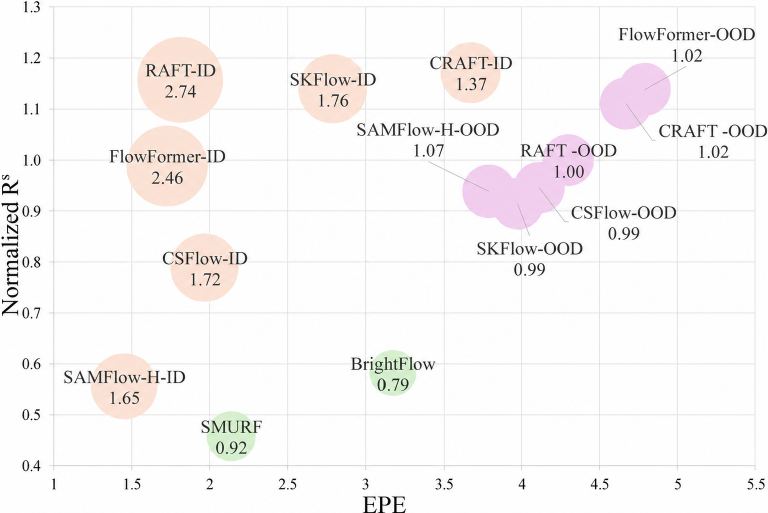}
    \caption{\textbf{Clean EPE, reference-normalized $R^{s}$, and $R^{s}_{\mathrm{rel}}$ on KITTI-FC.} Bubble size represents $R^{s}_{\mathrm{rel}}$, with the corresponding value shown below each model name.}
    \label{fig:kitti-fc-EPE-RCRE-RCREr}
\end{figure}

\noindent \textbf{Finding 4:} \textit{Clean accuracy is associated with, but does not determine, corruption robustness.}
Models with lower clean EPE often achieve lower unnormalized $R^{s}$ and $R^{d}$, but exceptions and the behavior of the relative metrics show that robustness cannot be inferred from clean accuracy alone.

\noindent \textbf{Finding 5:} \textit{Clean-domain fine-tuning does not guarantee proportional robustness.}
As illustrated in Fig.~\ref{fig:kitti-fc-EPE-RCRE-RCREr}, ID fine-tuning can improve EPE while $R^{s}_{\mathrm{rel}}$ increases. This indicates that improvements on clean in-domain samples need not transfer proportionally to corrupted inputs.
SEA-RAFT provides a further example: fine-tuning reduces its clean EPE from $3.8$ to $1.7$, yet its $R^{s}$ increases from $14.0$ to $43.6$ under Snow and from $2.6$ to $5.5$ under Camera Motion Blur (Tab.~\ref{table:kitti_rcre}).

\begin{table}[t]
\caption{\textbf{OOD stability of spatial modeling designs on KITTI-FC.} Category-average $R^{s}$ is normalized corruption-wise by RAFT-OOD (lower is better).}
\label{tab:ood_spatial_range}
\centering
{
\scriptsize
\setlength{\tabcolsep}{2pt}
\renewcommand{\arraystretch}{1.08}
\begin{tabular}{lcccccc}
\toprule
Model & Digital & Illum. & Weather & Noise & Blur & Avg. \\
\midrule
SKFlow     & \textbf{0.8298} & \textbf{0.7801} & \textbf{0.9034} & 1.1370 & 0.9657 & \textbf{0.9146} \\
GMFlowNet  & 0.9232 & 1.0818 & 0.9237 & \textbf{0.9129} & \textbf{0.8603} & 0.9378 \\
\bottomrule
\end{tabular}}

    \vskip -1.5\baselineskip plus -1fil
\end{table}

\noindent \textbf{Finding 6:} \textit{The beneficial spatial modeling range depends on the corruption mechanism.}
Relative to GMFlowNet, SKFlow has a smaller, bounded receptive field despite enlarging local context with super kernels~\cite{sun2022skflow}; GMFlowNet uses explicit global matching and therefore has a larger global matching range~\cite{zhao2022global}. As shown in Table~\ref{tab:ood_spatial_range}, SKFlow is more stable under Digital, Illumination, and Weather corruptions, whereas GMFlowNet performs better under Noise and Blur. This contrast is consistent with a mechanism-dependent benefit from their different spatial modeling ranges, but it is not a controlled architecture ablation. Neither design is uniformly optimal across corruption mechanisms.

\begin{figure}[!t]
    \centering
    \includegraphics[width=\linewidth]{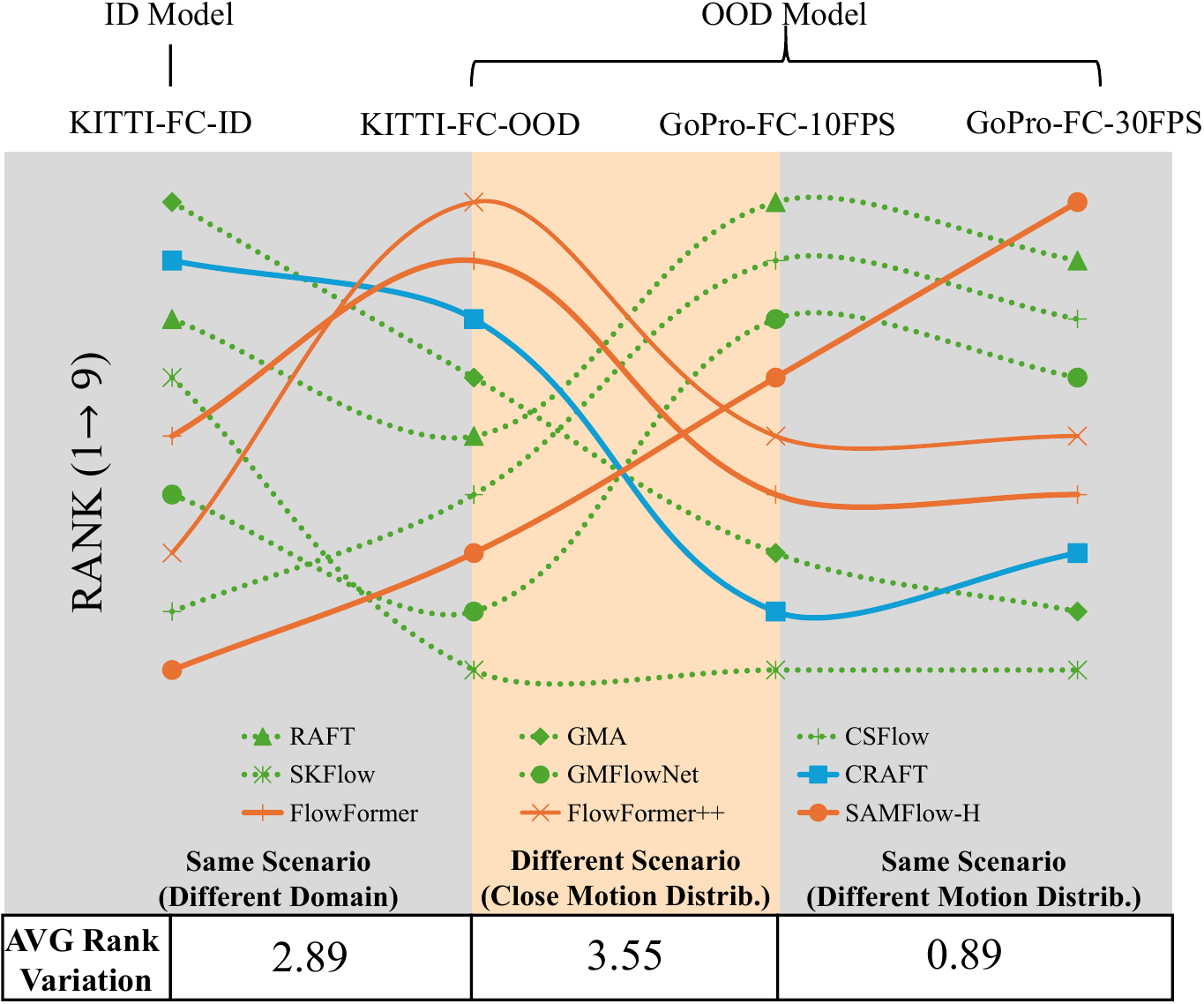}
    \caption{\textbf{Reference-normalized $R^{s}$ rank across benchmark settings.} Among the matched methods shown, none maintains consistent superiority across KITTI-FC ID/OOD and GoPro-FC 10/30 FPS.}
    \label{fig:rank across datasets}
    \vskip -1.5\baselineskip plus -1fil
\end{figure}

\noindent \textbf{Finding 7:} \textit{No evaluated method maintains a consistent rank across all settings.}
Figure~\ref{fig:rank across datasets} shows rank changes between KITTI-FC ID/OOD and GoPro-FC 10/30 FPS. For the five methods with matched KITTI-FC and GoPro-FC results, the overall KITTI--GoPro Spearman rank correlation is $0.30$ ($p=0.624$). This small matched set does not support a significance claim, but it demonstrates that the observed ordering is not stable enough to justify a universal winner. The smaller changes between GoPro sampling rates apply only to the tested displacements and may not generalize to extreme motion.

\section{Conclusion and Discussion}
\label{sec:conclusion}

This work introduces a spatiotemporal corruption benchmark based on captured KITTI and GoPro data, evaluating driving-domain and motion-distribution robustness through GT-free output stability $R^{s}$ and GT-based accuracy degradation $R^{d}$. Local structural loss is especially harmful under the evaluated corruption schedules, clean-domain fine-tuning does not guarantee proportional robustness, and method rankings vary across settings. These findings motivate robustness as a design objective beyond clean EPE. Benefits from global context and foundation-model features remain setting-dependent. Future methods should preserve local correspondence cues and temporal consistency under corruption.

\noindent\textbf{Limitations.}
KITTI-FC inherits the limited scale and annotation characteristics of KITTI-2015. Its 80-pair test split provides a controlled driving-domain evaluation but does not exhaust the diversity of real deployment conditions. GoPro-FC expands the diversity of captured egocentric motion, but it does not provide ground-truth optical flow. Consequently, it supports only the GT-free $R^{s}$ analysis and cannot support $R^{d}$, $R^{s}_{\mathrm{rel}}$, or $R^{d}_{\mathrm{rel}}$. Moreover, although the source images and videos are captured in real scenes, most corruptions are algorithmic approximations rather than recordings of adverse conditions. The benchmark should therefore be interpreted as a controlled diagnostic proxy that complements rendered and directly captured adverse-condition datasets.

\bibliographystyle{IEEEtran}
\bibliography{bib}

\section{Paired-frame controls on KITTI-FC}
\label{sec:supp_temporal}
We additionally compare cross-frame corruption settings for \textit{Over Exposure} and \textit{Snow} to assess how corruptions affect frame pairs. We evaluate RAFT-OOD on 80 KITTI-FC pairs across five levels with 12 iterations, reporting $R^s/R^d$ in pixels. For \textit{Over Exposure}, the benchmark's one-clean/one-corrupted pair is compared with applying the same exposure change to both frames; each affected frame retains the original severity. For \textit{Snow}, independently generated layers are compared with a shared image-fixed layer. This shared layer is a static-overlay control, not a simulation of physically coherent snow-particle trajectories.

\noindent\textbf{Paired-frame corruptions.}
Table~\ref{tab:r3_temporal} shows the comparison. For \textit{Over Exposure}, the degradation gap between asymmetric and equal two-frame exposure increases at higher severity, consistent with stronger inter-frame inconsistency in local information. For \textit{Snow}, a shared image-fixed layer may impair estimation through persistent occlusion, yielding larger errors than independent layers. These observations are consistent with Finding~2 on the importance of local correspondence information and support dedicated temporal-corruption evaluation. These controls compare cross-frame settings; they do not isolate a unique causal contribution of temporal inconsistency.

\begin{table}[htbp]
\centering
\caption{\textbf{Paired-frame controls on KITTI-FC.} Asymmetric exposure is more harmful at high severity, while shared snow is more harmful across all levels.}
\label{tab:r3_temporal}
\begingroup
\footnotesize
\setlength{\tabcolsep}{2pt}
\begin{tabular}{@{}ccccc@{}}
\toprule
& \multicolumn{2}{c}{\textit{Over Exposure}} & \multicolumn{2}{c}{\textit{Snow}} \\
Level & One frame & Both frames & Independent & Shared \\
\midrule
1 & 0.60/0.09 & 0.66/0.09 & 5.16/3.68 & 21.46/19.90 \\
2 & 1.59/0.77 & 1.63/0.74 & 11.77/9.63 & 26.48/24.77 \\
3 & 2.73/1.92 & 2.04/1.04 & 11.30/8.80 & 26.71/24.96 \\
4 & 4.66/3.69 & 2.88/1.62 & 18.33/15.24 & 29.32/27.57 \\
5 & 6.90/5.73 & 3.40/1.80 & 13.35/10.62 & 26.20/24.45 \\
\bottomrule
\end{tabular}

\endgroup
\end{table}

\FloatBarrier
\section{Motion-scale controls on GoPro-FC}
\label{sec:supp_motion}
For each frame rate, we sample 40 pairs from each of five videos, yielding 200 pairs per rate. RAFT-OOD evaluates the same pairs under JPEG and Snow at all five severity levels. Since GoPro provides no flow GT, mean clean-predicted flow magnitude is used as a model-dependent motion proxy. For pair $i$, with stability error $R^s_i$ and mean clean-predicted magnitude $\bar m_i$ in pixels, we compute
\[
Q=\frac{1}{N}\sum_{i=1}^{N}\frac{R^s_i}{\bar m_i+10^{-6}}.
\]
Here, $N=200$ per frame rate. We compute $Q$ at each severity and report the mean over all five levels in Table~\ref{tab:r3_motion}, alongside the five-level mean $R^s$. Normalization divides each pair's mean error by its mean clean-predicted motion magnitude. For the stratified comparison, we use predicted-motion ranges $[5,10)$, $[10,20)$, $[20,40)$, and $[40,80)$ px, each containing at least five pairs per frame rate. Common weights are proportional to the smaller sample count in each range, retaining 163/156 pairs at 10/30 FPS. The controls cover one model and two corruption types; sampled pairs within videos can be correlated.

\noindent\textbf{Motion scale on GoPro-FC.}
Table~\ref{tab:r3_motion} shows that larger absolute changes at 10 FPS coexist with smaller motion-relative changes. Comparing $R^s$ within common predicted-displacement ranges reduces the five-level JPEG gap from 1.812 to 0.625 px; the Snow gap reverses. These results are compatible with the original absolute comparison and echo the distinction between absolute and relative responses in Finding~5, although $Q$ uses motion magnitude rather than clean EPE. The model-dependent controls do not establish a universal frame-rate ordering.

\begin{table}[htbp]
\centering
\caption{\textbf{GoPro-FC motion-scale analysis.} Larger absolute prediction changes at 10 FPS coexist with smaller motion-normalized changes.}
\label{tab:r3_motion}
\begingroup
\footnotesize
\setlength{\tabcolsep}{4pt}
\begin{tabular}{@{}lcc@{}}
\toprule
Corruption & $R^s$ (10/30 FPS) & $Q$ (10/30 FPS) \\
\midrule
JPEG & 3.113/1.302 & 0.0952/0.1441 \\
Snow & 5.479/3.664 & 0.1593/0.2643 \\
\bottomrule
\end{tabular}

\endgroup
\end{table}

\FloatBarrier
\section{ID fine-tuning settings}
\label{sec:supp_id}
Supervised models are initialized from their corresponding OOD checkpoints and fine-tuned on the clean versions of the 120 KITTI-FC training pairs, with standard augmentations and no benchmark-corrupted inputs. The settings in Table~\ref{tab:id_finetuning} follow each method's KITTI fine-tuning recipe. We use AdamW with weight decay $10^{-5}$.

\begin{table}[!ht]
\centering
\caption{\textbf{Supervised ID fine-tuning settings.} All methods use clean KITTI-FC training pairs with standard augmentations.}
\label{tab:id_finetuning}
{\begingroup
\scriptsize
\setlength{\tabcolsep}{2pt}
\renewcommand{\arraystretch}{1.15}
\begin{tabularx}{\linewidth}{@{}Xrrrr@{}}
\toprule
Method & LR & Step & Batch & Crop ($H\!\times\!W$) \\
\midrule
RAFT, CSFlow, GMFlowNet, FlowDiffuser & $10^{-4}$ & 50k & 6 & $288\!\times\!960$ \\
GMA, CRAFT & $1.25\!\times\!10^{-4}$ & 30k & 6 & $288\!\times\!960$ \\
SKFlow & $1.75\!\times\!10^{-4}$ & 30k & 6 & $288\!\times\!960$ \\
FlowFormer, FlowFormer++ & $1.25\!\times\!10^{-4}$ & 50k & 6 & $432\!\times\!960$ \\
SAMFlow-T/B/H & $1.25\!\times\!10^{-4}$ & 50k & 6 & $400\!\times\!720$ \\
DPFlow & $10^{-4}$ & 50k & 3 & $288\!\times\!960$ \\
SEA-RAFT-M & $10^{-4}$ & 10k & 16 & $432\!\times\!960$ \\
\bottomrule
\end{tabularx}
\endgroup
}
\end{table}

\FloatBarrier
\section{Quality-controlled comparison for Finding 2}
\label{sec:supp_severity}
Finding~2 compares corruption responses under the benchmark's original severity schedules. To examine whether the observed trend persists at comparable image quality, we evaluate JPEG and Spatter against Saturate using RAFT-OOD on KITTI-FC. For each source pair, we consider all 25 combinations of the five severity levels and retain those with mean two-frame PSNR and SSIM gaps no larger than 1 dB and 0.03. Among eligible combinations, we select the one minimizing $|\Delta\mathrm{PSNR}|+20|\Delta\mathrm{SSIM}|$, independently of the flow results. Each source pair contributes at most once per comparison. PSNR uses full-resolution images, while SSIM uses images downsampled to one-quarter width and height.

\begin{table}[!ht]
\centering
\caption{\textbf{Quality-controlled comparison using RAFT-OOD.} JPEG and Spatter produce larger stability errors and accuracy degradation than Saturate at comparable image quality.}
\label{tab:r3_severity_supp}
\begingroup
\footnotesize
\setlength{\tabcolsep}{3pt}
\begin{tabular}{@{}lcc@{}}
\toprule
 & JPEG / Saturate & Spatter / Saturate \\
\midrule
Eligible pairs & 33 & 14 \\
Mean PSNR (dB) & 26.25 / 25.95 & 18.27 / 18.06 \\
Mean SSIM & 0.945 / 0.960 & 0.756 / 0.759 \\
Mean $R^s$ (px) & 8.00 / 1.37 & 21.73 / 2.42 \\
Mean $R^d$ (px) & 5.71 / 0.52 & 19.46 / 1.33 \\
\midrule
$\Delta R^s$ (px) & +6.63 & +19.31 \\
95\% CI & [4.28, 9.32] & [14.36, 24.26] \\
$\Delta R^d$ (px) & +5.19 & +18.13 \\
95\% CI & [2.90, 7.98] & [12.19, 24.05] \\
\bottomrule
\end{tabular}
\endgroup
\end{table}

Table~\ref{tab:r3_severity_supp} reports the mean image-quality and flow metrics for the retained samples. Slash-separated entries follow the corruption order in each column; differences subtract Saturate from JPEG or Spatter. Flow metrics use KITTI GT-valid pixels, and intervals are obtained from 3,000 paired bootstrap resamples.

Despite similar PSNR and SSIM, both JPEG and Spatter cause substantially larger prediction shifts and accuracy degradation than Saturate. Both contrasts have positive mean differences in $R^s$ and $R^d$, with all four confidence intervals above zero. The agreement between $R^s$ and $R^d$ shows that the larger output changes are accompanied by greater accuracy loss, rather than merely different predictions. This supports Finding~2's emphasis on local correspondence cues: comparable aggregate image quality need not imply comparable preservation of the information used for optical-flow estimation.

PSNR and SSIM provide approximate image-quality controls. These results support the observed trend for RAFT-OOD within the available quality overlap. Under the joint PSNR/SSIM criteria, JPEG--Fog has no eligible pairs and Spatter--Fog has only one, which is insufficient for a meaningful matched comparison.

\FloatBarrier
\section{Official-checkpoint comparison with RobustSpring}
\label{sec:supp_checkpoint}
We evaluate all 80 KITTI-FC test pairs using the official RAFT-Sintel checkpoint specified in RobustSpring's Appendix~A. Table~\ref{tab:r3_checkpoint} reports mean clean-to-corrupted prediction shifts over KITTI GT-valid pixels at 12 iterations and, as a sensitivity check, 32 iterations. RobustSpring values are taken from its published RAFT results; its complete inference settings are not aligned here. Mean full-resolution SSIM for JPEG, Spatter, and Glass Blur is 0.694, 0.731, and 0.729, respectively, compared with the reported 0.70, 0.72, and 0.73.

\begin{table}[!ht]
\centering
\caption{\textbf{Official RAFT-Sintel checkpoint comparison (prediction shift in pixels).} Larger shifts persist for JPEG and Spatter, while Glass Blur becomes comparable.}
\label{tab:r3_checkpoint}
\begingroup
\footnotesize
\setlength{\tabcolsep}{4pt}
\begin{tabular}{@{}lccc@{}}
\toprule
 & \multicolumn{2}{c}{KITTI-FC} & RobustSpring \\
\cmidrule(lr){2-3}
Corruption & 12 iter. & 32 iter. & Published \\
\midrule
JPEG, level 5 & 10.639 & 10.299 & 3.190 \\
Spatter, level 4 & 17.197 & 17.019 & 7.980 \\
Glass Blur, level 1 & 1.078 & 1.057 & 1.050 \\
\bottomrule
\end{tabular}

\endgroup
\end{table}

JPEG and Spatter retain larger prediction shifts with the official checkpoint, whereas Glass Blur becomes comparable to the published result. These observations support a descriptive cross-benchmark comparison; differences in datasets, corruptions, and evaluation protocols preclude attributing the shifts solely to captured-scene content. The official Sintel training recipe includes KITTI data, so this control is separate from our KITTI-unseen RAFT-OOD evaluation.

\end{document}